\documentclass[jcp,preprint]{revtex4-1}

\usepackage{amsmath,amssymb}
\usepackage{graphicx,color}

\usepackage[latin1]{inputenc}

\newcommand{\ket}[1]{|#1\rangle}
\newcommand{\bra}[1]{\langle#1|}
\newcommand{\braket}[1]{\langle#1\rangle}
\newcommand{\rmd}{\mathrm{d}}
\newcommand{\rmi}{\mathrm{i}}
\renewcommand{\Im}{\operatorname{Im}}
\renewcommand{\Re}{\operatorname{Re}}
\newcommand{\Tr}{\operatorname{Tr}}

\newcommand{\pdiff}[2]{\frac{\partial #1}{\partial #2}}

\newcommand{\boldpsi}{\boldsymbol{\psi}}

\newcommand{\vphi}{\varphi}
\newcommand{\tphi}{\tilde{\phi}}
\newcommand{\tPhi}{\tilde{\Phi}}
\newcommand{\tPsi}{\tilde{\Psi}}
\newcommand{\tvphi}{\tilde{\varphi}}
\newcommand{\larrow}{\longrightarrow}
\newcommand{\tc}{\tilde{c}}

\begin{document}

\title{\emph{Ab initio} quantum dynamics using coupled-cluster}

\author{Simen Kvaal}
\email{simen.kvaal@cma.uio.no}
\affiliation{University of Oslo, Centre of Mathematics for Applications,
  N-0316 Oslo, Norway}
\affiliation{University of Oslo, Centre for Theoretical and
  Computational Chemistry,
  N-0315 Oslo, Norway}

%\keywords{ab initio \sep quantum dynamics \sep coupled-cluster}

%\email{simen.kvaal@cma.uio.no}
%\affiliation{Matematisches Institut, Universität Tübingen, Auf der
%  Morgenstelle 10, D-72076 Tübingen, Germany}
%\affiliation{Centre of Mathematics for Applications, University of
%  Oslo, N-0316 Oslo, Norway} 

%\pacs{xxx}

\begin{abstract}
  The curse of dimensionality (COD) limits the current
  state-of-the-art {\it ab initio} propagation methods for
  non-relativistic quantum mechanics to relatively few particles. For
  stationary structure calculations, the coupled-cluster (CC) method
  overcomes the COD in the sense that the method scales polynomially
  with the number of particles while still being size-consistent and
  extensive. We generalize the CC method to the time domain while
  allowing the single-particle functions to vary in an adaptive
  fashion as well, thereby creating a highly flexible, polynomially
  scaling approximation to the time-dependent Schrödinger
  equation. The method inherits size-consistency and extensivity from
   the CC method. The method is dubbed orbital-adaptive
  time-dependent coupled-cluster (OATDCC), and is a hierarchy of
  approximations to the now standard multi-configurational
  time-dependent Hartree method for fermions. A numerical experiment
  is also given.
\end{abstract}

\maketitle

%\tableofcontents

\section{Introduction}
\label{sec:introduction}

% Angle of article: the dynamics people. Needs quite a lot of CC
% material.

Presently, the most advanced \emph{ab initio} approximations to the
time-dependent Schrödinger equation for a system of identical
particles are the multiconfigurational time-dependent Hartree methods
for fermions (MCTDHF) and variants
\cite{Meyer2009,Beck2000,Meyer1990}. These methods apply the
time-dependent variational principle
\cite{Broeckhove1988,Kramer1981,Lubich2005} to an $N$-body
wavefunction ansatz being a Slater determinant expansion using a
finite (incomplete) set of $L\geq N$ orbitals $\vphi_p$ with creation
operators $c^\dag_p$ (with $\{c_p,c^\dag_q\} = \delta_{pq}$ for
fermions),
\begin{equation*}
  \ket{\Psi_\text{MCTDHF}} \equiv \sum_{p_1}\sum_{p_2>p_1}\cdots\sum_{p_N>p_{N-1}}
  A_{p_1\cdots p_N} c^\dag_{p_1} c^\dag_{p_2} \cdots c^\dag_{p_N} \ket{-},
\end{equation*}
where both the amplitudes $A_{p_1\cdots p_N}$ and the orbitals
$\vphi_p$ are free to vary in single-particle space. Varying the
orbitals in this way is especially important if studies of unbound
systems are desired, such as the study of ionization of atoms or
molecules. The key point is that, if the orbitals are not optimized, a
system in the continuum would need a \emph{huge} fixed
basis. Moreover, the \emph{variational} determination of the orbitals
compresses the wavefunction in a quasi-optimal way: from time $t$ to
$t+dt$, the basis is changed as so to minimize the $L^2$ norm error of
the wavefunction \cite{Lubich2005}.

While powerful, MCTDHF still suffers from exponential scaling of
computational complexity with respect to the number of particles $N$
present; the somewhat prosaically termed ``curse of dimensionality''
(COD). The effect of the variational determination of the
single-particle functions can be said to be a postponing of the COD to
higher particle numbers. For example, for the simple time-dependent
Hartree--Fock method (TDHF), i.e., MCTDHF using precisely $L=N$
orbitals and therefore only a \emph{single} determinant $\ket{\phi}$,
qualitatively good results may be achieved, even if the system is
unbound \cite{Zanghellini2004}. 

One may attempt at reducing the exponential scaling by truncating the
Slater determinant expansion at, say, single and double excitations
relative to one of the determinants $\ket{\phi} = c^\dag_1\cdots
c^\dag_N\ket{-}$, considered as a ``reference determinant'',
\begin{equation}
  \ket{\Psi_\text{MCTDHF-SD}} = \left(1 + \sum_{ia} A_i^a c^\dag_a c_i
  \ket{\phi} +  \frac{1}{2!^2} \sum_{ijab} A_{ij}^{ab} c^\dag_a c_i
  c^\dag_b c_j \right)\ket{\phi} \equiv (1 + A_{SD})\ket{\phi},
\label{eq:intro-mctdhf-sd-ansatz}
\end{equation}
hoping that the higher-order excited determinants' contribution can be
neglected.  (We have arbitrarily chosen intermediate normalization
$\braket{\phi|\Psi}=1$. In the sums, $i,j\leq N$ and $a,b>N$ is
assumed.) This would achieve polynomial scaling but would destroy the
important property of size-consistency \cite{Helgaker2002,Shavitt2009}: approximation of
non-interacting subsystems separately at the singles and doubles level
would not be consistent with approximating the whole at the same
level; \emph{independent} excitations of the subsystems are neglected,
giving rise to artificial correlation effects.

In this article, we develop a time-dependent version of the popular
coupled-cluster (CC) method for fermions, where we allow the orbitals
to vary in a similar fashion to MCTDHF. We call the method
\emph{orbital adaptive time-dependent coupled-cluster}
(OATDCC). Formally, the two methods are very similar, with closely
related equations of motion.  The main difference is the fact
that coupled-cluster is not variational in the usual sense, rather, it
is naturally cast in a \emph{bivariational} setting, a generalization
of the variational approach \cite{Arponen1983}. Bivariational
functionals are \emph{complex analytic}, while the standard
variational functionals are manifestly real. Moreover, approximations of
\emph{both} the wavefunction $\ket{\Psi}$ the complex conjugate
$\bra{\Psi}$ must be introduced. For the version of CC that has now become
standard (referred to as ``standard CC'' in this paper), the
wavefunctions in the singles and double approximation (CCSD) are
parametrized according to
\begin{subequations}
  \label{eq:intro-cc-ansatz}
  \begin{align}
  \ket{\Psi_\text{CC}} &\equiv e^T\ket{\phi}, \quad T = \sum_{ai} \tau_i^a
  c^\dag_a \tilde{c}_i + \frac{1}{2!^2} \sum_{ijab} \tau_{ij}^{ab} c^\dag_a
  \tilde{c}_i c^\dag_b \tilde{c}_j + \cdots \label{eq:intro-cc-ansatz-1} \\
  \bra{\tPsi_\text{CC}} &\equiv \bra{\tphi} (1 + \Lambda) e^{-T}, \quad \Lambda =
  \sum_{ai} \lambda_a^i c^\dag_i \tc_a + \frac{1}{2!^2}\sum_{ijab}
  \lambda_{ab}^{ij} c^\dag_i \tc_a c^\dag_j \tc_b + \cdots \label{eq:intro-cc-ansatz-2}
  \end{align}
\end{subequations}
where $\bra{\tPsi_\text{CC}}$ approximates $\bra{\Psi}/\braket{\Psi|\Psi}$.
The operator $T$ is called a \emph{cluster operator} and produces
excitations with respect to a reference Slater determinant $\ket{\phi}$. The
amplitudes $\tau_i^a$ and $\tau_{ij}^{ab}$ correspond (to first order) to the expansion
coefficients $A_i^a$ and $A_{ij}^{ab}$ of Eqn.~(\ref{eq:intro-mctdhf-sd-ansatz}). The
cluster operator $\Lambda$ is a de-excitation operator, and its
amplitudes $\lambda_a^i$ and $\lambda_{ab}^{ij}$ are essentially
the parameters of $\bra{\tPsi_\text{CC}}$, which also is composed of
excitations relative to a reference bra determinant $\bra{\tphi}$.
To anticipate the developments in later sections, we have introduced creation and
annihilation operators with respect to \emph{biorthogonal orbitals}
$\vphi_p$ and $\tvphi_q$, i.e.,
\begin{equation*}
  \left\{ \tilde{c}_p, c^\dag_q \right\} = \tilde{c}_p c^\dag_q + c^\dag_q
  \tilde{c}_p \equiv \braket{\tilde{\vphi}_p|\vphi_q},
\end{equation*}
That is to say, $\ket{\Psi}$ is built using the $\vphi_p$, while $\bra{\tPsi}$ is
built using $\tvphi_p$. This relaxation of orthonormality of the
orbitals is necessary to ensure that the bivariational functional is
complex analytic if the orbitals are to be treated as variational
parameters, as discussed in Section~\ref{sec:bi-vp-3}.

Our treatment can be viewed as a generalization of the
standard CC Lagrangian approach to linear response theory
\cite{Dalgaard1983,Arponen1983,Helgaker1989,Koch1990}, where the
$\Lambda$ amplitudes are time-dependent Lagrangian multipliers
as introduced in a constrained
minimization of the CC energy. However, we emphasize the bivariational point of
view, where $\Lambda$ becomes a part of the wave function
parametrization. In Ref.~\cite{Pedersen2001} even biorthogonal
orbitals were consideres.

For an excellent introduction to CC theory, see the article
\cite{Crawford2000} by Crawford, and the review \cite{Bartlett2007} by
Bartlett and Musia\l, as well as the textbooks \cite{Shavitt2009} by
Shavitt and Bartlett and \cite{Harris1992} by Harris {\it et al}. For
the present work, the article \cite{Arponen1983} is fundamental, as it
casts the CC theory in the bivariational framework, an approach not
emphasized in most introductions to CC theory.

The OATDCC method considers the standard CC ansatz plugged into the
proper bivariational functional, see Section~\ref{sec:oatdcc-functional}
below. In addition to having the $T$ and $\Lambda$ amplitudes as
degrees of freedom, the orbitals $\ket{\vphi_p}$ and approximations
$\bra{\tvphi_p}$ to their complex conjugates are varied. It turns out
the relaxation of the orbitals makes the singles amplitudes $\tau_i^a$
and $\lambda_a^i$ redundant. Truncation of the remaining terms at
doubles, triples, etc, then gives a hierarchy of approximations with
TDHF at one end ($T=\Lambda=0$), and full MCTDHF (no truncation of $T$
or $\Lambda$) at the other. Inbetween we have the doubles
approximation (OATDCCD), doubles-and-triples approximation (OATDCCDT),
and so on. Considering the success of the CC method for structure
calculations and the MCTDHF method for dynamics, OATDCC should be a
viable alternative to MCTDHF with asymptotically much lower cost but good
accuracy. Importantly, OATDCC is size-consistent.

There are few applications of CC methods to \emph{ab initio} dynamics
in the literature. It was, however, proposed as early as 1978 by
Schönhammer and Gunnarsson \cite{Schonhammer1978}, and independently by
Hoodbhoy and Negele \cite{Hoodbhoy1978,Hoodbhoy1979}, who even
considered time-dependent orbitals, albeit with an \emph{explicit}
time-dependence. We shall discuss their approach briefly in
Section \ref{sec:method-relations}. Recently, the standard CC ansatz
using a fixed basis was applied to laser-driven dynamics of some small
molecules \cite{Huber2011}, but with expectation values calculated in
a different way from the usual CC approach.

This article is written with the MCTDHF community in mind. Since
bivariational principles are rarely considered (which is true for the
CC community as well), we give a somewhat detailed discussion in
Section~\ref{sec:bi-vp}. As CC theory might be unfamiliar, and as we
use a somewhat unfamiliar variational approach to derive CC theory,
Section~\ref{sec:cc-theory} is devoted to the basics of the CC
formalism, leading up to the case where the orbitals are varied
freely. In Section~\ref{sec:oatdcc} we discuss the OATDCC functional
and derive the corresponding equations of motion. We then 
perform a simple numerical experiment in Section~\ref{sec:experiment}
to demonstrate the method before we conclude the paper.

CC calculations invariably involve a lot of algebra. Therefore, in
\ref{sec:algebraic}, algebraic expressions for various quantities
appearing in the OATDCCD method are listed. These are generated using
symbolic algebra software developed with the \textsc{SymPy} library
for the programming language \textsc{Python} \cite{SymPy2009}. An
independent derivation of MCTDHF is given in
\ref{sec:mctdhf-derivation} in order to shed further light on
the connections between the variational and the bivariational
principles. It may also serve as a helpful device for the researchers
in the audience not familiar with MCTDHF theory.

No attempt is made to be mathematically rigorous in this article, as
like the standard CC method
\cite{Schneider2009,Rohwedder2011,Rohwedder2011a} and the MCTDHF
method \cite{Lubich2005,Koch2007b,Conte2010,Bardos2010} such an
analysis is expected to be quite involved.  Instead, we make formal
computations as if all operators present were bounded or the spaces
finite dimensional.

\section{Bivariational principles}
\label{sec:bi-vp}

\subsection{Functionals}
\label{sec:bi-vp-2}

Let $A$ be an operator over Hilbert space
$\mathcal{H}$, and consider the functional
\begin{equation*}
  \mathcal{E}_A : \mathcal{H}' \times \mathcal{H} \longrightarrow
  \mathbb{C}, \quad  \mathcal{E}_A[\bra{\Psi'}, \ket{\Psi}] \equiv
  \frac{\braket{\Psi'|A|\Psi}}{\braket{\Psi'|\Psi}}
\end{equation*}
defined whenever the expression makes sense. Note that the arguments
are \emph{two independent} Hilbert space elements. The functional
$\mathcal{E}_A$ is a  generalization of the expectation value
functional to operators that are not necessarily Hermitian, and for
obvious reasons it may be called \emph{the bivariational expectation
  value functional} \cite{Arponen1983,Lowdin1989}.  Consider the conditions for
vanishing first variation, $\delta\mathcal{E}_A = 0$, for all
independent variations of $\bra{\Psi'}$ and $\ket{\Psi}$. A
straightforward formal calculation gives the stationary conditions
\begin{equation}
  (A - a)\ket{\Psi} = 0 \quad \text{and} \quad \bra{\Psi'}(A - a) = 0,
  \label{eq:evp}
\end{equation}
with
\begin{equation*}
  a = \mathcal{E}_A[\bra{\Psi'},\ket{\Psi}]
\end{equation*}
being the value of $\mathcal{E}_A$ at the critical point. In other
words $\bra{\Psi'}$ and $\ket{\Psi}$ with $\braket{\Psi'|\Psi}\neq
0$ are left- and right eigenvectors,
respectively, of the operator $A$, with eigenvalue $a$.  Computing the eigenvalues and
eigenvectors from $\delta\mathcal{E}_A=0$ is therefore referred to as ``the
bivariational principle.''

Since the system Hamiltonian $H=H^\dag$ is the generator for the time-evolution of
the system, we consider the following bivariational generalization of
the usual action functional \cite{Arponen1983,Kramer1981}, i.e.,
\begin{align}
  \mathcal{S}[\bra{\Psi'},\ket{\Psi}] &\equiv \int_0^T
  \frac{\braket{\Psi'(t)| \left(\rmi\hbar\pdiff{}{t} -
        H\right)|\Psi(t)}}{\braket{\Psi'(t)|\Psi(t)}}  \; \rmd t
  \notag \\ &= 
\int_0^T
  \rmi\hbar \frac{\braket{\Psi'(t)|\pdiff{}{t}\Psi(t)}}{\braket{\Psi'(t)|\Psi(t)}}
  - \mathcal{E}_H[\bra{\Psi'(t)},\ket{\Psi(t)}] \; \rmd t, \label{eq:td-functional}
\end{align}
where it is understood that the functional depends on the whole
history of the system from time $t=0$ to $t=T$. Suppose $\mathcal{S}$
is stationary ($\delta\mathcal{S}=0$) under all variations of
$\bra{\Psi'}$ and $\ket{\Psi}$ vanishing at the endpoints
$t=0,T$. Straightforward manipulations now give, up to irrelevant
time-dependent phase constants,
\begin{equation}
  \rmi\hbar \pdiff{}{t}\ket{\Psi(t)} = H\ket{\Psi(t)}
  \quad\text{and}\quad -\rmi\hbar \pdiff{}{t}\bra{\Psi'(t)} =
  \bra{\Psi'(t)} H.
  \label{eq:se-from-bivar}
\end{equation}
Consequently, we note that the time-dependent Schrödinger equation and its complex
conjugate arise from a \emph{time-dependent bivariational
  principle}. 

In both the stationary and time-dependent case, it is convenient to do
a reparametrization $\bra{\tPsi}=\braket{\Psi'|\Psi}^{-1}\bra{\Psi'}$
so that
\begin{equation}
  \braket{\tPsi|\Psi} = 1,
  \label{eq:normalization}
\end{equation} 
eliminating the denominator in each
functional. For the stationary case, this effectively eliminates one
of the two eigenvalue equations (\ref{eq:evp}), and makes
$\bra{\tPsi}$ a unique function of $\ket{\Psi}$ and vice versa. To see
this, note that since (a)
eigenvectors corresponding to different eigenvalues are \emph{always}
orthogonal, and since (b) $\braket{\tPsi|\Psi}=1$, $\bra{\tPsi}$ is
uniquely given by $\ket{\Psi}$ at the critical point as the
biorthogonal left eigenvector corresponding to the right eigenvector
$\ket{\Psi}$. 

For the time-dependent case, the normalization
$\braket{\tPsi(t)|\Psi(t)} = 1$ eliminates one of the Schrödinger
equations (\ref{eq:se-from-bivar}), but the phase ambiguity is still
present for the remaining equation, a similar situation as the stationary case.

Unless otherwise stated, the normalization (\ref{eq:normalization}) is
assumed in the following, and it is indicated by the tilde
$\bra{\tPsi}$ instead of the prime $\bra{\Psi'}$.

It is also natural to assume some secondary
normalization on $\ket{\Psi}$, such that the critical point actually
becomes locally unique for both the time-dependent and
time-independent cases. (``Locally unique'' means that there may be many
critical points but that they are isolated.) This is done in CC
theory, where $\braket{\tphi|\Psi}=1$ is assumed. This removes phase
ambiguity in both time-dependent and time-independent pictures.

\subsection{Generating approximations from submanifolds}
\label{sec:bi-vp-3}

Contrary to the usual variational principles, in the bivariational
principles \emph{the wavefunction and its complex conjugate are
  formally independent}.  When applied to Hermitian operators this
opens up possibilities for more general approximations compared to the
standard variational principles to spectra and dynamics, when the
variations are restricted to some predefined approximation manifold.
However, where the usual ``Hermitian'' time-dependent variational
principle is, to paraphrase Kramer and Saraceno \cite{Kramer1981}, a
deaf and dumb procedure that always gives an answer, the bivariational
principle requires a more careful approach, as we will discuss in this
section.

Perhaps for this reason, the bivariational principles are little
known. The author has found only a few relevant sources in the literature
apart from Arponen's seminal coupled-cluster paper \cite{Arponen1983},
the most relevant ones being a brief mention by Killingbeck in his
review on perturbation theory \cite{Killingbeck1977} and a discussion
by Löwdin {\it et al.}~\cite{Lowdin1989} concerning self-consistent
field-calculations on non-Hermitian (complex scaled) Hamiltonians. The
bivariational principle seems largely unexplored.

It is important to note that the standard critique of coupled-cluster
is that it is ``non-variational''. While it is true that it makes
estimation of errors harder, it is not  a serious drawback
in any other sense, since the calculation is firmly rooted in a
variational principle. For example, if $H$ does not
depend explicitly on time, $(d/dt) \mathcal{E}_H \equiv 0$, i.e.,
energy is conserved. Probability is always conserved,
$(d/dt)\mathcal{E}_1 = (d/dt) \braket{\tPsi|\Psi} \equiv 0$. Just like
the usual time-dependent variational principle, these are simple
consequences of the symmetries of the action functional.

For bivariational approximations, one introduces different
parametrizations of the wavefunction $\ket{\Psi}$ and its complex
conjugate $\bra{\Psi}$, which is contrary to the usual variational
principle. Notice that in Eqn.~(\ref{eq:intro-cc-ansatz}), {different}
parametrizations $\ket{\Psi_\text{CC}}$ and $\bra{\tPsi_\text{CC}}$
are used, but $\braket{\tPsi_\text{CC}|\Psi_\text{CC}}=1$. Formally,
we do a variation over a manifold $\mathcal{M}\subset
\mathcal{H}'\times\mathcal{H}$, i.e.,
$(\bra{\tPsi},\ket{\Psi})\in\mathcal{M}$. Alternatively, one may think
of $\mathcal{M}$ as a subset of rank-one density operators $u =
\ket{\tPsi}\bra{\Psi}$, $\Tr(u)=1$, where $u^\dag \neq u$ is allowed.

As $\mathcal{H}$ is a complex space, the functionals $\mathcal{E}_H$
and $\mathcal{S}$ are complex. On the other hand, in the usual
variational principle, the functionals are always real-valued. This
has some interesting consequences. We now briefly discuss four
important aspects: the analytic structure of the functionals, the need
for systematic refinability of $\mathcal{M}$, interpretations of
complex critical points, and even-dimensionality of $\mathcal{M}$.

The parametrization of $u=(\bra{\tPsi},\ket{\Psi})$ must be complex
analytic, at least locally. A parametrization which is \emph{not}
analytic is, in essence, a real parametrization, since it depends on
both the real and imaginary parts of the (local) coordinates
$z\in\mathbb{C}^n$ separately, and not only $\Re z + \rmi \Im z$. Thus, we have
$2n$ real coordinates. Since \emph{both} $\Re\delta\mathcal{S}$ and
$\Im\delta\mathcal{S}$ must vanish, this leads to $4n$
equations. Unless there is some extra structure, i.e., that
$\Im\mathcal{S}\equiv 0$ such as in the standard variational
principle, a solution cannot be expected to exist. Correspondingly,
the bivariational functionals should be complex analytic. Nowhere
should explicitly real parameters occur, and nowhere should parameters
be explicitly complex conjugated.

Suppose the system Hamiltonian $H$ is bounded from below, i.e., the
expectation value is bounded from below. This is the source of the
usefulness of the Hermitian stationary variational principle, since \emph{any}
parametrization gives an upper bound for the ground state energy. A
potential danger with the bivariational expectation value functional
is that it is \emph{not} bounded from below, even if $H$ is. Indeed, $\mathcal{E}_H$
is complex analytic and can take on values in the whole of
$\mathbb{C}$. One cannot insert ``just anything'' and hope to get
sensible results by computing critical points. To avoid this problem,
and to allow for the computation of error estimates, $\mathcal{M}$
should be chosen in a way that is in some sense systematically refinable
towards the full space $\mathcal{H}' \times \mathcal{H}$, e.g.,
there is some discretization parameter that can be used to measure the
accuracy. In CC theory, this parameter is the truncation level of the
cluster operators and the number $L$ of orbitals used.

Critical values of $\mathcal{E}_H$ for $H=H^\dag$ may be complex, even
though the exact eigenvalues are always real. However,
if $H=H^\dag$, the imaginary values of the critical values 
generally ``should be small'' if $\mathcal{M}$ is chosen ``well enough'', and
may correspondingly be ignored in order to assign a physical interpretation to
the critical value, i.e., energy. This is also justified by the fact that the
functional $\Re\mathcal{E}_H$ has the same critical points as
$\mathcal{E}_H$ if the parametrization is analytic.

For the approximate manifold $\mathcal{M}$, we must be certain that
the critical point $(\bra{\tPsi},\ket{\Psi})$ is locally unique. In
particular, this is necessary for the corresponding critical
\emph{value}, i.e., $\mathcal{E}_A[\bra{\tPsi}\ket{\Psi}]$, to be unique
for any observable $A$. Otherwise, the physical state cannot be said
to be well-defined. Intuitively, the parameters must then come in pairs; roughly
stated every parameter in $\ket{\Psi}$ should have a parameter in
$\bra{\tPsi}$.

This can be shown explicitly.  Suppose
$\bra{\tPsi}$ and $\ket{\Psi}$ are parametrized locally using some
set of complex variables $z(t) = \mathbb{C}^n$, i.e., we have an
approximating manifold
$\mathcal{M}\subset\mathcal{H}'\times\mathcal{H}$ whose dimension
is assumed to be finite for simplicity.  When inserted into the
time-dependent bivariational functional, we obtain a new functional
$F[z(\cdot)] = \mathcal{S}[\bra{\tPsi(z(\cdot))},\ket{\Psi(z(\cdot))}]$, whose
stationary point is given by the solution of the differential equation
\begin{equation}
  \rmi\hbar C(z) \dot{z} = \nabla_z E(z), \quad E(z) =
  \mathcal{E}_H[\bra{\tPsi(z)},\ket{\Psi(z)}]. 
  \label{eq:local-coords}
\end{equation}
The matrix $C(z)$ is given by
\begin{equation*}
  C(z)_{jk} = \Bigg\langle{\pdiff{\tPsi}{z_j}\Bigg|\pdiff{\Psi}{z_k}}\Bigg\rangle -
  \Bigg\langle{\pdiff{\tPsi}{z_k}\Bigg|\pdiff{\Psi}{z_j}}\Bigg\rangle, 
\end{equation*}
which is complex anti-symmetric. For any anti-symmetric matrix, if
$\lambda$ is an eigenvalue of multiplicity $m$, so is $-\lambda$, implying that $C(z)$ is
not invertible if $n$ is odd, since it must have a zero
eigenvalue. Consequently, the approximation manifold must be complex even
dimensional if Equation~(\ref{eq:local-coords}) is to have a unique
solution.

\section{Coupled-cluster functionals}
\label{sec:cc-theory}

\subsection{Biorthogonality of orbitals}

The fact that we require our bivariational functionals to be analytic
is important, as it necessitates the relaxation of the orthonormality
the orbitals $\varphi_p$ by introducing a second set of biorthogonal
orbitals $\tilde{\varphi}_q$, being in effect \emph{approximate}
complex conjugates of each other. However, they are introduced as
\emph{independent} complex parameters. 

Consider a subspace $\mathcal{V}\subset\mathcal{H}$, generated by a
finite set of orbitals $\Phi = (\varphi_1,
\varphi_2,\cdots,\varphi_L)$, which we write $\mathcal{V} =
\mathcal{V}[\Phi]$. (We interpret $\vphi_p$ as the $p$th column of the
matrix $\Phi$.) These orbitals need not be orthonormal; it is only the
one-body space spanned by the $\varphi_p$ that matters. In the
bivariational functionals, we desire to vary $\bra{\tPsi}$ in as large
space $\tilde{\mathcal{V}}\subset\mathcal{H}^\dag$ as possible, while
guaranteeing the existence of a dual vector non-orthogonal to
$\ket{\Psi}\in\mathcal{V}$. (Otherwise, the denominator in
Eqn.~(\ref{eq:td-functional}) may vanish.) The only restriction on the
space $\tilde{\mathcal{V}}$ is that it is generated by a set of dual
orbitals $\tilde{\Phi} = (\tilde{\varphi}_1; \cdots;
\tilde{\varphi}_L)$ (where we interpret $\tvphi_p$ as the $p$th
\emph{row} of $\tPhi$). Sometimes we will stress the fact that
$\vphi_p$ and $\tvphi_p$ are single-particle ket and bra-functions,
respectively, by explicitly writing $\ket{\vphi_p}$ and
$\bra{\tvphi_p}$.

Consider the overlap matrix $S$ with matrix elements $S_{pq} =
\braket{\tvphi_p|\vphi_q}$.  Since the spaces $\mathcal{V}[\Phi]$ and
$\tilde{\mathcal{V}}[\tPhi]$ only depend on the subspaces spanned by
$\Phi$ and $\tPhi$, respectively, we may via a suitable transform
assume that $S$ is diagonal with only $1$s and $0$s on the
diagonal. If $S$ is invertible, then the orbitals are biorthogonal,
\begin{equation*}
  \braket{\tilde{\varphi}_p|\varphi_q} = \delta_{pq}.
\end{equation*}

It is straightforward to show the following claim: the existence of a
$\bra{\tPsi}\in\tilde{\mathcal{V}}[\tPhi]$ for every
$\ket{\Psi}\in\mathcal{V}[\Phi]$ such that $\braket{\tPsi|\Psi}\neq 0$
is equivalent to requiring the overlap matrix $S_{pq} =
\braket{\tvphi_p|\vphi_q}$ to be invertible, i.e., that the orbitals
are biorthogonal. (A corresponding claim where the roles of
$\ket{\Psi}$ and $\bra{\tPsi}$ are reversed is equivalent.)

Biorthogonality is equivalent to
\begin{equation*}
  \braket{\tilde{\phi}_{p_1\cdots p_N}|\phi_{q_1,\cdots,q_N}} =
  \delta_{p_1,q_1}\cdots\delta_{p_N,q_N}
\end{equation*}
for the Slater determinant bases of $\tilde{\mathcal{V}}$ and
$\mathcal{V}$, given by
\begin{equation*}
  \ket{\phi_{p_1\cdots p_N}} \equiv c^\dag_{p_1} c^\dag_{p_2}\cdots
  c^\dag_{p_N} \ket{-} \quad\text{and}\quad
  \bra{\tphi_{q_1\cdots q_N}} \equiv \bra{-} \tc_{q_N} \tc_{q_{N-1}}
  \cdots \tc_{q_1},
\end{equation*}
respectively. (We assume $p_1<p_2<\cdots$ and $q_1<q_2<\cdots$.) 

To prove the claim, suppose that $\braket{\tvphi_{p'}|\vphi_{p'}} =
0$.  Then clearly, if $\ket{\Psi} = \ket{\phi_{p',p_2\cdots,p_N}}$, no
$\bra{\Psi'}\in\tilde{\mathcal{V}}$ is non-orthogonal to
$\ket{\Psi}$. Conversely, suppose that no such $p'$ exists. Then,
given an arbitrary $0\neq \ket{\Psi}\in\mathcal{V}$,
$\braket{\tilde{\phi}_{p_1,\cdots,p_N}|\Psi}$ must be nonzero for some
$p_1,\cdots,p_N$. This proves the claim.

The
creation operators are defined using field creation and annihilation
operators as
\begin{subequations}
\begin{align}
  c^\dag_p &\equiv \int \varphi_p(x) \boldpsi(x)^\dag \rmd x
  \label{eq:creation-def} \intertext{and}
  \tilde{c}_p &\equiv \int \tilde{\varphi}_p(x) \boldpsi(x) \rmd x.
  \label{eq:annihilation-def}
\end{align}
\end{subequations}
The biorthogonality condition implies an anticommutator
relation
\begin{equation}
  \left\{ \tilde{c}_p, c^\dag_q \right\} \equiv \tilde{c}_p c^\dag_q +
  c^\dag_q \tilde{c}_p \equiv \braket{\tilde{\varphi}_p|\varphi_q} =
  \delta_{pq},
  \label{eq:bi-anti-comm}
\end{equation}
proven by inserting the definitions~(\ref{eq:creation-def}) and
(\ref{eq:annihilation-def}). Thus, Wick's theorem
\cite{Wick1950,Shavitt2009} holds in its usual form, simply replacing $c_p$
with the operator $\tc_p$.

In standard CC theory, and virtually every other manybody method,
$\varphi_p(x)^* \equiv \tilde{\varphi}_p(x)$, so that $\mathcal{V}
\equiv \tilde{\mathcal{V}}^\dag$. We stress that the relaxation of
this requirement allows for a complex analytic functional, which is
essential for the bivariational principle.

\subsection{Excitation operators}
\label{sec:exponential-ansatz}

The orbitals are divided into \emph{occupied} (the $N$ first) and
\emph{virtual} orbitals (the $L-N$ last). By common convention,
indices $i$, $j$, $k$ etc.~denote occupied orbitals, while indices
$a$, $b$, $c$, etc.~denote virtual orbitals. The terminology comes
from the fact that CC can be considered a perturbational scheme, where
the exact wavefunction is written as
\begin{equation}
  \ket{\Psi} = e^T \ket{\phi}, \label{eq:exp-ansatz}
\end{equation}
where $\ket{\phi}$ is a reference zeroth order approximation Slater
determinant 
\begin{equation*}
  \ket{\phi} = c^\dag_1 c^\dag_2 \cdots c^\dag_N \ket{-},
\end{equation*}
and $T$ is an operator on the form
\begin{equation*}
  T = \sum_{ai} \tau_i^a c^\dag_a \tilde{c}_i +
  \frac{1}{2!^2} \sum_{abij} \tau_{ij}^{ab} c^\dag_a \tilde{c}_i c^\dag_b \tilde{c}_j
  + \cdots.
\end{equation*}
The operator $c^\dag_a \tilde{c}_i$ destroys a particle in an occupied orbital
(creates a hole) and creates a ``virtual'' particle above the Fermi
sea defined by $\ket{\phi}$. The first sum on the right-hand-side is a \emph{singles
  excitation operator}, while the second sum is a \emph{doubles
  excitation operator}, and so on. A general
$n$-fold excitation operator is on the form
\begin{equation*}
  T_n = \frac{1}{n!^2} \sum_{i_1\cdots i_n} \sum_{a_1\cdots a_n}
  \tau_{i_1\cdots i_n}^{a_1\cdots a_n} c^\dag_{a_1} \tilde{c}_{i_1} \cdots
  c^\dag_{a_n} \tilde{c}_{i_n}.
\end{equation*}
It is easy to see that without loss of generality the amplitudes can
be taken to be anti-symmetric, which is the reason for the
combinatorial prefactor. Importantly, since the occupied and virtual
orbitals are disjoint sets, \emph{all} excitation operators commute.
It is convenient to introduce a generic index $\mu$ for the
excitations, i.e.,
\begin{equation*}
  T = \sum_\mu \tau^\mu X_\mu,
\end{equation*}
where $X_\mu$ is a shorthand for $X_{a}^{i} = c^\dag_a \tilde{c}_i$,
$X_{ab}^{ij} = X_a^i X_b^j$, and so on. Note that in the latter
expansion, only linearly independent excitations are included,
eliminating the combinatorial factors. 

It is worthwhile to note, that as an operator, $T$ depends on
\emph{both} the amplitudes $\tau = (\tau^\mu)$ \emph{and} the orbitals
through the $X_\mu$, that is, $T = T(\tau,\tPhi,\Phi)$. It standard CC
theory, one usually thinks of $T$ as the primary unknown, since the
orbitals are fixed, and since dependence on $\tau$ is linear and
one-to-one. In the OATDCC theory we must be careful, for example when
computing $\partial T/\partial t$. Usually, however, there should be
no danger of confusion when, for brevity, we suppress the parameter dependence of
excitation operators.

We also note, that even though $X_\mu$ depends
explicitly on the dual orbitals through the appearance of
$\tilde{c}_i$, the function $\ket{\phi_\mu} \equiv X_\mu\ket{\phi}$
does not: $\tilde{c}_i$ is only responsible for removing $\varphi_i$
(not $\tvphi_i$!) from a determinant. The Slater determinants $\ket{\phi_\mu}$ are easily
seen to form a basis for $\mathcal{V}$ together with $\ket{\phi}$.

We define de-excitation operators as operators on the form
\begin{equation*}
  S = \sum_\mu \sigma_\mu \tilde{X}^\mu = \sum_{ai} \sigma_a^i c^\dag_i \tilde{c}_a +
  \frac{1}{2!^2} \sum_{abij} \sigma_{ab}^{ij} c^\dag_i \tilde{c}_a c^\dag_j \tilde{c}_b
  + \cdots.
\end{equation*}
The term ``de-excitation operator'' has an
obvious interpretation, but note however that these operators \emph{excite} bra
determinants. In particular,
\begin{equation*}
  \braket{\tilde{\phi}^\mu|\phi_\nu} =
  \braket{\tilde{\phi}|\tilde{X}^\mu X_\nu|\phi} = \delta_\nu^\mu.
\end{equation*}
Correspondingly, excitation operators de-excite bra states.

We now comment on the form of the Hamiltonian in second
quantization. For simplicity, we assume that the Hamiltonian contains
at most two-body forces, i.e.,
\begin{equation*}
  H = \sum_{i=1}^N h(i) + \frac{1}{2}\sum_{i,j, i\neq j} u(i,j),
\end{equation*}
in first quantization,
where $h(i)$ is an operator acting only on the degrees of freedom for
particle $i$, and $u(i,j)$ acts only on the degrees of freedom of the
pair $(i,j)$. For molecular electronic systems in the Born--Oppenheimer
approximation, $h(i)$ is the sum of kinetic energy and the nuclear
attraction potential, while $u(i,j)$ is the Coulomb repulsion between
electrons $i$ and $j$. Suppose now $\Pi$ is the projection operator
\begin{equation*}
  \Pi \equiv \ket{\phi}\bra{\tphi} + \sum_\mu \ket{\phi_\mu}\bra{\tilde{\phi}^\mu},
\end{equation*}
which acts as identity on $\tilde{\mathcal{V}}[\tPhi]$ and $\mathcal{V}[\Phi]$:
For any $\ket{\Psi}\in\mathcal{V}$, $\Pi\ket{\Psi} =
\ket{\Psi}$, and for any $\bra{\Psi'}\in\tilde{\mathcal{V}}$,
$\bra{\Psi'}\Pi = \bra{\Psi'}$. However, $\Pi\neq\Pi^\dag$ so it is
not an orthogonal projector. We now have
\begin{equation*}
  \braket{\Psi'|H|\Psi} = \braket{\Psi'|\Pi H \Pi|\Psi},
\end{equation*}
where
\begin{align}
  \Pi H \Pi &= \sum_{pq} \braket{\tilde{\varphi}_p|h|\varphi_q}
  c^\dag_p \tilde{c}_q + \frac{1}{4} \sum_{prqs}
  \braket{\tilde{\varphi}_p \tilde{\varphi}_r | u | \varphi_q
    \varphi_s }_\text{AS} c^\dag_p c^\dag_r \tilde{c}_s \tilde{c}_q
  \notag \\
 &\equiv \sum_{pq} h^p_q c^\dag_p \tc_q + \frac{1}{4}\sum_{prqs} u^{pr}_{qs}
 c^\dag_p c^\dag_r \tc_s \tc_q. \label{eq:projected-h}
\end{align}
The two-body integrals are
anti-symmetrized according to the standard in CC theory and are given by
\begin{align}
  \braket{\tilde{\varphi}_p \tilde{\varphi}_r | u | \varphi_q
    \varphi_s }_\text{AS} &\equiv \braket{\tilde{\varphi}_p
    \tilde{\varphi}_r | u | \varphi_q \varphi_s } -
  \braket{\tilde{\varphi}_p \tilde{\varphi}_r | u | \varphi_s
    \varphi_q },\label{eq:two-body-integral-1} \\
  \braket{\tilde{\varphi}_p \tilde{\varphi}_r | u | \varphi_q
    \varphi_s } &\equiv \int
  \tvphi_p(x)\tvphi_r(y)u(x,y)\vphi_q(x)\vphi_s(y) \;\rmd x\rmd
  y \label{eq:two-body-integral-2}
\end{align}
It is important to
note, that unless the orbitals are complete, $\Pi H \Pi \neq H$.

Similar considerations as the above also hold for arbitrary one- and
two-body operators.

\subsection{From variational to bivariational CC}
\label{sec:cc-standard}

Having discussed orbitals and operator expressions using second
quantization, we now turn to the CC ansatz.
It is a fundamental fact of CC theory that
\emph{any} wavefunction $\ket{\Psi}\in\mathcal{V}$ normalized
according to $\braket{\tphi|\Psi}=1$ can be written on the form
\eqref{eq:exp-ansatz}, i.e., the exponential ansatz is covers the
whole of the discrete Hilbert space $\mathcal{V}$. To see this, we
simply observe that $\exp(T) = 1 + A$, where $A$ is a new excitation
operator. By writing $T = T_1 + T_2 + \cdots$ and $A = A_1 + A_2 +
\cdots$, $A$ and $T$ can be compared term-by term, giving explicit
formulae for $T_k$ in terms of $A_\ell$, $\ell\leq k$, and vice
versa. Since any $\ket{\Psi}\in\mathcal{V}$ with $\braket{\tphi|\Psi}=1$
can be written $\ket{\Psi} = \ket{\phi} + A\ket{\Phi}$, the result
follows. 

Similarly, we have $\bra{\Psi'} = \bra{\tilde{\phi}} e^{T'}$ for any
$\bra{\Psi'}$ normalized according to $\braket{\Psi'|\phi}=1$, where
\begin{equation*}
  T' = \sum_\mu (\tau')^\mu \tilde{X}_\mu
\end{equation*}
is a de-excitation operator. Note that the exponential parametrization
results hold for any choice of biorthogonal orbitals.

We note that a truncation in $A$ instead of $T$ at, say, $A=A_1+A_2$ gives a
\emph{linear} parametrization which defines the CI singles and doubles
ansatz, CISD. [Compare also with
Eqn.~(\ref{eq:intro-mctdhf-sd-ansatz}).]

The bivariational expectation $\mathcal{E}_H$ now reads
\begin{equation}
  \mathcal{E}_H[\tau',\tau,\tPhi,\Phi] =
  \frac{\braket{\tphi|e^{T'}He^T|\phi}}{\braket{\tphi|e^{T'}e^T|\phi}},
  \label{eq:bivar-step-1}
\end{equation}
where we note that the functional dependence on the orbitals is
implicit in the reference determinants and $T$ and $T'$. 

We now make the
observation, that $\braket{\omega|\phi}=1$, where
\begin{equation*}
  \bra{\omega} \equiv \left(\braket{\tphi|e^{T'}e^T|\phi}\right)^{-1}\bra{\tphi}e^{T'}e^T,
\end{equation*}
implying that there exists a de-excitation operator $S=\sum_\mu\sigma_\mu\tilde{X}^\mu$ such that $\bra{\omega} =
\bra{\tphi} e^S$. From this we obtain
\begin{equation*}
  \bra{\Psi'} = \braket{\tphi|e^{T'}e^T|\phi}\bra{\omega}e^{-T} =
 \braket{\tphi|e^{T'}e^T|\phi} \bra{\tphi}e^S e^{-T} \equiv  \braket{\tphi|e^{T'}e^T|\phi}\bra{\tPsi} .
\end{equation*}
Inserting this into (\ref{eq:bivar-step-1}), we get rid of the
denominator, viz,
\begin{equation}
  \mathcal{E}_H[\sigma, \tau,\tPhi,\Phi] = \braket{\tPsi|H|\Psi} =
  \braket{\tphi|e^{S}e^{-T}He^T|\phi}. 
\label{eq:bivar-step-2}
\end{equation}
We stress that there is no loss of generalization in these
manipulations.

We perform a further change of variables. Writing $e^S = I + S + \cdots
\equiv I + \Lambda$ with $\Lambda=\sum_\mu \lambda_\mu\tilde{X}^\mu$,
partially transforming from exponential to linear parametrization of
$\bra{\tPsi}$,
\begin{equation}
  \mathcal{E}_H[\lambda, \tau,\tPhi,\Phi] = \braket{\tphi|(I +
    \Lambda)e^{-T}He^T|\phi}.
  \label{eq:bivar-step-3}
\end{equation}
Disregarding the dependence on $\tPhi$ and $\Phi$, this
functional is the well-known CC expectation functional
\cite{Arponen1983,Helgaker1989,Koch1990}. The usual interpretation for
$\lambda_\mu$ is as Lagrange multipliers for a constrained
minimization of the energy which is equivalent to the standard CC
equations \cite{Crawford2000}. In our case, however, it is interpreted
as part of the parametrization of the approximate dual wavefunction
which enters the expectation functional. $\lambda_\mu$ should be
treated on equal footing with $\tau^\mu$: they are equally important.

All three functionals (\ref{eq:bivar-step-1}),
(\ref{eq:bivar-step-2}), and (\ref{eq:bivar-step-3}) are equivalent to
the multi-configurational Hartree--Fock (MCHF) functional. 
The fundamental approximation idea in CC theory is now to \emph{truncate}
the expansion for $T$ and $\Lambda$ (or $S$ if the functional
(\ref{eq:bivar-step-2}) is used) at a finite excitation level. For example, for
the coupled-cluster singles and doubles (CCSD) approximation, one
takes
\begin{equation*}
  T \approx T_1 + T_2 \quad \text{and} \quad \Lambda \approx \Lambda_1 + \Lambda_2,
\end{equation*}
neglecting amplitudes with $n>2$. In the following discussion, the
truncation level $n$ should be considered a parameter of the
ansatz.  Under such a truncation, the three CC functionals
$\mathcal{E}_H[\tau',\tau,\tPhi,\Phi]$, $\mathcal{E}_H[\sigma,\tau,\tPhi,\Phi]$ and
$\mathcal{E}_H[\lambda,\tau,\tPhi,\Phi]$ are no longer equivalent.

Suppose for the moment that the orbitals are held fixed and orthonormal. What methods
do the three functionals
(\ref{eq:bivar-step-1})--(\ref{eq:bivar-step-3}) define? The
functional (\ref{eq:bivar-step-1}) corresponds to a variational
coupled-cluster theory (VCC) within the chosen basis and is abandoned
for reasons that will soon be apparent. The functional
(\ref{eq:bivar-step-2}) defines the so-called extended CC (ECC)
functional \cite{Arponen1983}, and defines an alternative approach to
the standard CC defined by the CC Lagrangian
(\ref{eq:bivar-step-3}). In fact, it may be viewed as a more
``natural'' version of CC since $\bra{\tPsi}$ is parametrized
exponentially, i.e., size-consistently.

In the remainder of the paper, we focus on the standard CC Lagrangian
(\ref{eq:bivar-step-3}), but with the orbitals are included as
variational parameters. We correspondingly refer to the functional as
the ``OATDCC expectation functional'' in the rest of the paper.

For the reader not acquainted with CC theory, it may be hard to see
that we have actually simplified matters with the CC Lagrangian or
ECC functional compared to simply considering the VCC expectation
value functional, with $T' = T^\dag$, which would be the default
approach of optimizing the energy using any ansatz. The problem is
that while the ansatz scales polynomially, the evaluation of
$\bra{\phi}e^{T^\dag}H e^T\ket{\phi}/\braket{\phi|e^{T^\dag}e^T|\phi}$
does not. It also seems hard to find size-extensive approximations of
finite order in $T$ \cite{Kutzelnigg1991}.

On the other hand, one of the
basic observations of CC is that Baker--Campbell--Hausdorff (BCH)
expansion of the similarity transform $\exp(-T)H\exp(T)$ truncates
\emph{identically} after a finite number of terms, regardless of the
number of particles present. In fact, for a Hamiltonian with at most two-body
potentials only terms up to fourth order are nonzero,
\begin{equation*}
  e^{-T}H e^T = H + \sum_{n=1}^4 \frac{1}{n!} [H,T]_n,
\end{equation*}
where $[H,T]_{n} = [[H,T]_{n-1},T]$ is the $n$-fold nested
commutator. This no less than remarkable fact can be seen from
$[[c^\dag_p,X_\mu],X_\nu]=[[\tilde{c}_p,X_\mu],X_\nu]=0$ (verified by
direct computation) and by the fact that $H$ is a fourth order
polynomial in the creation and annihilation operators.

It follows that $\mathcal{E}_H[\Lambda,T,\tPhi,\Phi]$ is a fourth
order polynomial in $\tau=(\tau^\mu)$ and linear in
$\lambda=(\lambda_\mu)$. This polynomial can be evaluated using
Wick's theorem. As this involves an agonizing amount of algebra, an
alternative approach resides in the use of Feynman graphs to simplify
Wick's theorem \cite{Crawford2000,Shavitt2009}, or in the use of
computer algebra software \cite{Hirata2006}. The latter approach has
become more common in the last years and is utilized here, see
\ref{sec:algebraic}.

\subsection{The OATDCC action functional}
\label{sec:oatdcc-functional}

Having established the form of the OATDCC expectation
functional, we now turn to the evaluation of the corresponding
action-like functional
$\mathcal{S}$ defining the Schrödinger dynamics,
\begin{equation*}
  \mathcal{S}[\lambda,\tau,\tilde{\Phi},\Phi] = \int_0^T
  \braket{\tilde{\phi}|(1 + \Lambda)e^{-T} (\rmi\hbar \pdiff{}{t} -
    H)e^T|\phi} \; \rmd t.
\end{equation*}
To evaluate the explicit functional dependence on the time derivatives
of $\tau$ and $\Phi$,  we must compute
$\pdiff{}{t}\ket{\Psi}=\pdiff{}{t}e^T\ket{\phi}$. To this end, we use the expansion
\begin{equation*}
  \ket{\Psi} = \ket{\phi} + \sum_\mu A^\mu \ket{\phi_\mu}, \quad A^\mu
  = A^\mu(\tau)
  = \braket{\tilde{\phi}^\mu|e^T|\phi},
\end{equation*}
where the summation is \emph{not} truncated at any level.
Since Wick's theorem only uses the anti-commutator
(\ref{eq:bi-anti-comm}), the coefficients $A^\mu=A^\mu(\tau)$ do not depend
explicitly on the orbitals, only on the (possibly truncated) amplitudes
$\tau$. Moreover, we compute the derivative of a Slater
determinant via
\begin{align*}
  \pdiff{}{t} c^\dag_{p_1} c^\dag_{p_2} \cdots c^\dag_{p_N} \ket{-} &=
  \dot{c}^\dag_{p_1} c^\dag_{p_2} \cdots c^\dag_{p_N} \ket{-}  +
  c^\dag_{p_1} \dot{c}^\dag_{p_2} \cdots c^\dag_{p_N} \ket{-} + \ldots
  \\
  &= \left(\sum_q \dot{c}^\dag_q \tilde{c}_q\right) c^\dag_{p_1}
  c^\dag_{p_2} \cdots c^\dag_{p_N} \ket{-}. 
\end{align*}
This implies
\begin{align}
  \pdiff{}{t}\ket{\Psi} &= \sum_\mu
  \left(\pdiff{}{t}A^\mu(\tau)\right)\ket{\phi_\mu} + \left(\sum_q \dot{c}^\dag_q
    \tilde{c}_q\right)\ket{\phi} + \sum_\mu
  A^\mu(\tau) \left(\sum_q \dot{c}^\dag_q
    \tilde{c}_q\right)\ket{\phi_\mu}  \notag \\
 &=  \left[\sum_\nu \dot{\tau}^\nu \pdiff{}{\tau^\nu} +  \left(\sum_q \dot{c}^\dag_q
   \tilde{c}_q\right)
 \right]\ket{\Psi} \notag
\\
 &=  \left(\sum_\nu \dot{\tau}^\nu X_\nu +  D
 \right)\ket{\Psi}, \quad D \equiv \left(\sum_q \dot{c}^\dag_q
   \tilde{c}_q\right). \label{eq:D-def-and-psi-derivative}
\end{align}
In the last step we used
\begin{equation*}
  \pdiff{}{\tau^\nu} \ket{\Psi} = X_\nu \ket{\Psi}.
\end{equation*}
For the time-derivative part of the functional integrand, we now get
\begin{align*}
  \rmi\hbar \braket{\tilde{\phi}|(1+\Lambda)e^{-T}\pdiff{}{t} e^T| \phi}
  &= \rmi\hbar \braket{\tilde{\phi}|(1+\sum_\mu\lambda_\mu
    \tilde{X}^\mu)e^{-T}\left(\sum_\nu \dot{\tau}^\nu X_\nu + D\right) e^T
    |\phi} \notag \\ &= \rmi\hbar\sum_\mu \lambda_\mu\dot{\tau}^\mu +
  \rmi\hbar\braket{\tilde{\phi}|(1+\Lambda)e^{-T}\Pi D \Pi e^T|\phi}.
\end{align*}
The projected operator $\Pi D \Pi$ is easily computed using
\begin{equation*}
  \ket{\dot{\varphi}_p} = (P+Q) \ket{\dot{\varphi}_p} = \sum_q
  \ket{\varphi_q}\braket{\tilde{\varphi}_q|\dot{\varphi}_p} + Q \ket{\dot{\varphi}_p}
\end{equation*}
where $P=\Phi\tPhi$ is the (oblique) projector onto single-particle
space, and $Q = 1-P$. It follows that
\begin{equation*}
 \Pi D \Pi =  \sum_{pq} \braket{\tilde{\varphi}_p|\dot{\varphi}_q} c^\dag_p
  \tilde{c}_q \equiv D_0.
\end{equation*}
Finally, we obtain
\begin{subequations}
  \label{eq:the-functional}
\begin{align}
  \mathcal{S}[\lambda,\tau,\tPhi,\Phi] &= \int_0^T \rmi \hbar \sum_\mu \lambda_\mu
  \dot{\tau}^\mu - \mathcal{E}_{H-\rmi\hbar  D_0}[\lambda,\tau,\tPhi,\Phi] \; \rmd t   \label{eq:the-functional-1}
\\
  &= \int_0^T \rmi\hbar \lambda_\mu\dot{\tau}^\mu +
   \rho^q_p (h^p_q - \rmi\hbar\eta^p_q) + \frac{1}{4} \rho^{qs}_{pr}
  u^{pr}_{qs} \; \rmd t   \label{eq:the-functional-2}
\end{align}
\end{subequations}
where
\begin{align*}
  \rho^q_{p} &= \rho^q_{p}(\lambda,\tau) \equiv
  \braket{\tilde{\phi}|(1+\Lambda) e^{-T} c^\dag_p \tilde{c}_q e^T
    |\phi}, \\
  \rho^{qs}_{pr} &= \rho^{qs}_{pr}(\lambda,\tau) \equiv
  \braket{\tilde{\phi}|(1+\Lambda) e^{-T} c^\dag_p c^\dag_r
    \tilde{c}_s \tilde{c}_q e^T
    |\phi}, \\
  h^p_q &= h_{pq}(\tPhi,\Phi) \equiv \braket{\tilde{\varphi}_p|
    h | \varphi_q}, \\
  \eta^p_q &= \eta^p_q(\tPhi,\dot{\Phi}) \equiv \braket{\tilde{\varphi}_p| \dot{\varphi}_q}, \\
  \intertext{and} u^{pr}_{qs} &= u^{pr}_{qs}(\tPhi,\Phi) \equiv
  \braket{\tilde{\varphi}_p \tilde{\varphi}_r | u | \varphi_q
    \varphi_s }_\text{AS}.
\end{align*}
In Eqn.~(\ref{eq:the-functional-2}) we introduced the Einstein
summation convention over repeated indices of opposite vertical
placement. This greatly simplifies the algebraic manipulation of CC
expressions, see \ref{sec:algebraic}. (Strictly speaking, we then should
introduce a similar index placement on the orbitals, e.g., $\tvphi^p$
instead of $\tvphi_p$, and $\tc^p$ instead of $\tc_p$. However, we
find this a too great departure from the standard, so we,
keep a somewhat inconsistent notation for simplicity.)

The quantities $\rho^q_p$ and $\rho^{qs}_{pr}$ (whose index placement
should be noted) are the CC reduced one-
and two-particle density matrices, respectively, and are not
explicitly dependent on the orbitals, since they are evaluated using
Wick's theorem depending only the fundamental anti-commutator. They
therefore only depend on the amplitudes. Similarly, the one-particle integrals
$h^p_q$, $\eta^p_q$ and the two-particle integrals $u^{pr}_{qs}$ only
depend on the orbitals. These facts dramatically simplify the application of the
evaluation of $\delta\mathcal{S}$.

\subsection{Standard CC and linear response}
\label{sec:standard-cc}

For clarity, we briefly discuss the standard CC method for the ground
state problem and the time-dependent generalization, i.e., the
equations of motion used in linear response theory \cite{Koch1990}.
For standard CC, a fixed set of orthonormal orbitals are chosen, i.e.,
\begin{equation*}
  \bra{\tvphi_p} \equiv \ket{\vphi_p}^\dag, \quad \left\{ c_p, c^\dag_q
  \right\} = \delta_{pq}.
\end{equation*}
The orbitals are usually but not necessarily chosen to be the
Hartree--Fock orbitals for the given (bound) system.

Since the orbitals are fixed, the only parameters in the expectation
value functional are the amplitudes $\tau$ and $\lambda$,
\begin{align}
  \mathcal{E}[\lambda,\tau] &= \braket{\phi|(1 + \Lambda)e^{-T}H
    e^T|\phi} \notag \\
  &= E_\text{CC}[\tau] + \sum_\mu \lambda_\mu
  \braket{\phi_\mu|e^{-T}He^T|\phi}, \quad E_\text{CC}[\tau] \equiv
  \braket{\phi|He^T|\phi},
  \label{eq:standard-cc-functional}
\end{align}
where we have used $\bra{\phi}T = 0$. Computing the derivatives with
respect to $\lambda_\mu$ and equating to zero, we get the stationary conditions
\begin{equation*}
  \braket{\phi_\mu|e^{-T}He^T|\phi} = 0, \quad \forall
  \mu\in\mathcal{I},
\end{equation*}
where the set $\mathcal{I}$ contains all the amplitudes in the desired
approximation, say CCSD. Note that we could arrive at this equation by simply
similarity-transforming the Schrödinger equation,
\begin{equation*}
  e^{-T}He^T\ket{\phi} = E \ket{\phi},
\end{equation*}
and projecting against $\bra{\phi_\mu}$. 

By projection onto $\bra{\phi}$ we get
\begin{equation*}
  E = \braket{\phi|He^T|\phi} = E_\text{CC}[\tau],
\end{equation*}
which is of course \emph{exact} within $\mathcal{V}=\tilde{\mathcal{V}}$ if the amplitudes are not
truncated. Truncations  give an
approximate similarity transformed Schrödinger equation, and the
interpretation of $\lambda_\mu$ as Lagrange multipliers for the
constrained minimization of $E_\text{CC}$ is then apparent.

In the CC literature, the coefficients $\lambda_\mu$ were originally
introduced in order to compute expectation values consistent with the
Hellmann--Feynman theorem
\cite{Monkhorst1977,Dalgaard1983,Helgaker1989,Koch1990}, i.e., that
one seeks an expectation value functional \emph{defined} by the
criterion
\begin{equation*}
  \braket{A} \equiv \pdiff{}{\epsilon}E_\text{CC}(\epsilon)\Big|_{\epsilon=0},
\end{equation*}
where $E_\text{CC}(\epsilon)$ is the energy eigenvalue approximation found with the CC
equations for the perturbed Hamiltonian $H
+ \epsilon A$. (Recall that $\braket{A}$ is the first order
perturbation in Rayleigh--Schrödinger perturbation theory.) It turns out, that $\braket{A} \equiv
\mathcal{E}_A[\lambda,\tau]$, where $(\lambda,\tau)$ is the critical point
of the functional (\ref{eq:standard-cc-functional}), i.e., the
unperturbed CC solution.

For the time-dependent case, one simply similarity transforms the
time-dependent Schrödinger equation, obtaining
\begin{equation*}
  \rmi\hbar\dot{T} \ket{\phi} = e^{-T}H e^{T}\ket{\phi},
\end{equation*}
and via projection,
\begin{equation*}
  \rmi\hbar\dot{\tau}^\mu = \bra{\phi_\mu}e^{-T}H e^{T}\ket{\phi}.
\end{equation*}
This is exact (within the space $\mathcal{V}$) for the untruncated
ansatz, but motivates the use of this equation in the truncated
case as well. In that case, it is easily obtained from the stationary
conditions of the CC Lagrangian
\begin{equation*}
  \mathcal{S}[\lambda,\tau] = \int_0^T
  \braket{\phi|(1+\Lambda)e^{-T}(\rmi\hbar \pdiff{}{t} - H)e^T|\phi}
  \; \rmd t = \int_0^T \rmi\hbar \lambda_\mu\dot{\tau}^\mu -
  \mathcal{E}[\lambda,\tau] \; \rmd t.
\end{equation*}
These are
\begin{subequations}
  \label{eq:response-eom}
\begin{align}
  \rmi\hbar\dot{\tau}^\mu &= \pdiff{}{\lambda_\mu}
  \mathcal{E}[\lambda,\tau] = \braket{\phi_\mu|e^{-T}He^T|\phi} \\
  -\rmi\hbar\dot{\lambda}_\mu &= \pdiff{}{\tau^\mu}
  \mathcal{E}[\lambda,\tau] =
  \braket{\phi|(1+\Lambda)e^{-T}[H,X_\mu]e^T|\phi},
\end{align}
\end{subequations}
as is easily verified; see \ref{sec:technical}. Expectation
values are computed using $\mathcal{E}_A[\lambda,\tau]$ as in the
stationary case.

\section{The OATDCC equations of motion}
\label{sec:oatdcc}

\subsection{Parametric redundancy}

In order to derive the equations of motion for $\tau = (\tau^\mu)$,
$\lambda = (\lambda_\mu)$, $\Phi=(\vphi_p)$ and $\tPhi=(\tvphi_p)$
given by the stationary condition $\delta \mathcal{S}=0$, we must, in
the functional $\mathcal{S}$, insert all \emph{independent} variations
of the parameters. However, for a given CC wavefunction pair $u =
(\bra{\tPsi},\ket{\Psi})\in\mathcal{M}$, there are many choices of the
amplitudes and orbitals that give the same pair. Correspondingly, not
all variations give independent equations, and this ambiguity must be
eliminated.

For example, it is well-known in CC theory that occupied and virtual
orbitals may be rotated among themselves with a corresponding inverse
rotation of the amplitudes, such that the wavefunction is actually
invariant to within a constant factor. 

To formalize this somewhat,
we write the collected parameters as a point $z$ in a manifold
$\mathcal{N}$, i.e., $z=(\lambda,\tau,\tPhi,\Phi)\in\mathcal{N}$. The
CC ansatz is then given by a many-to-one mapping $f(z)$,
\begin{equation*}
  f : \mathcal{N} \larrow \mathcal{M},
\end{equation*}
inducing functionals
\begin{equation*}
  E[z] \equiv \mathcal{E}_H[f(z)], \quad\text{and}\quad S[z(\cdot)] \equiv
  \mathcal{S}[f(z(\cdot))],
\end{equation*}
where we for the action functional explicitly have written out the dependence
on the history for clarity. Suppose we identify a Lie group
$\mathsf{G}$ such that
\begin{subequations}
  \label{eq:invariance}
\begin{align}
  E[G\circ z] = E[z] \; \text{for all $G
    \in \mathsf{G}$}, \; z\in\mathcal{N} \\
  S[G(\cdot)\circ z(\cdot)] = S[z(\cdot)] + C \; \text{for all $G(t)
    \in \mathsf{G}$}, \; z(t)\in\mathcal{N}
\end{align}
\end{subequations}
with $C$ being a constant (only dependent on
$G$). Equations~(\ref{eq:invariance}) states that the CC functionals
are invariant under the action of the Lie group. Let $z(t)$ be
given. We now observe, that there is a \emph{continuum} of histories
$G(t)\circ z(t)$, one for every choice of $G(\cdot)$, that corresponds
to the \emph{same} integrand. So any change $\delta z(t) = G(t)z(t)$
with $G(t)$ infinitesimal will \emph{not} change the value of the
action functional. Such variations of $z(t)$ will lead to redundant
equations. We comment that infinitesimal $G(t)$ is on the form $1 +
g(t)$, where $g(t)$ is in the Lie algebra of $\mathsf{G}$.

\begin{figure}
  \begin{center}
    \includegraphics{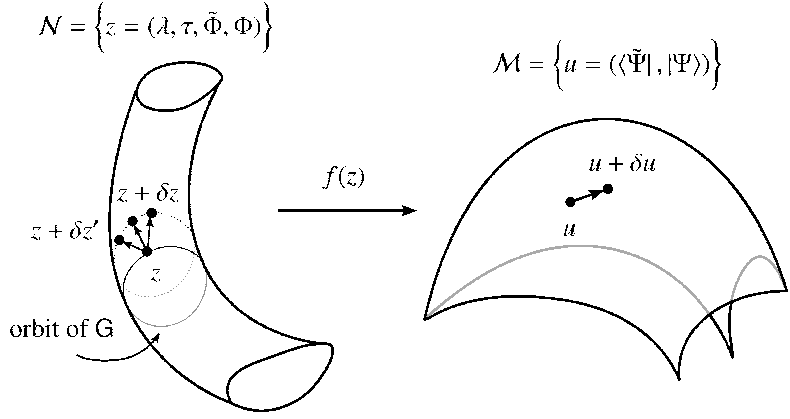}
    
    \caption{Illustration of the parameter manifold $\mathcal{N}$ and
      the CC wavefunction manifold $\mathcal{M}$. An element $u\in
      \mathcal{M}$ is given as the image $u = f(z)$, albeit in a
      non-unique way. The pre-image $f^{-1}(x)$ is the orbit of a Lie group
      $\mathsf{G}$ acting on some $z$, illustrated as a closed curves on $\mathcal{N}$
      passing through $z$. Thus, there is a manifold of displacements $\delta z$
      giving rise to the \emph{same} displacement $\delta
      u$. Specifying a gauge selects a unique $\delta z$ (a unique
      $G$) for each
      $\delta u$.\label{fig:manifolds}}
  \end{center}
\end{figure}

From the invariance of the functionals, the physics predicted by
$G\circ z$ is the same as that predicted by $z$. The physical solution
is unique, but the parameters are not. This situation is similar to
the one in gauge field theories, so the elimination of these extra
degrees of freedom in the parameters is called a gauge choice. In
Figure~\ref{fig:manifolds} the relationship between $z$ and $u=f(z)$
is illustrated, along with the concept of the invariance under the
action of $\mathsf{G}$.

The group $\mathsf{G}$ describing mixing of occupied and
virtual orbitals separately consists of all block diagonal and
invertible matrices on the form
\begin{equation*}
  G = 
  \begin{pmatrix}
    [G_{ij}] & 0 \\
    0 & [G_{ab}]
  \end{pmatrix},
\end{equation*}
i.e, the matrix elements $G_{ia} = G_{ai} = 0$. The action on $z$ is
defined as follows: The orbitals are transformed according to
\begin{equation}
  \Phi \larrow  \Phi G, \quad \tilde{\Phi} \larrow
   G^{-1}\tilde{\Phi}
  \label{eq:rotation},
\end{equation}
which preserves biorthogonality. Equation~(\ref{eq:rotation}) is
equivalent to the transformation
\begin{equation*}
  c^\dag_p \larrow \sum_q c^\dag_q G_{qp}, \quad   \tilde{c}_p \larrow
  \sum_q G^{-1}_{pq} \tilde{c}_q.
\end{equation*}
This gives the
following transformation on $X_a^i$:
\begin{equation*}
  X_a^i \larrow \sum_{bj} G^{-1}_{ij} X_b^j G_{ba}.
\end{equation*}
We define the transformation of the amplitudes by
\begin{subequations}
  \label{eq:amptrafo}
\begin{align}
  \tau_i^a &\larrow \sum_{i'a'} G_{ii'} \tau_{i'}^{a'} G^{-1}_{a'a} \\
  \tau_{ij}^{ab} &\larrow \sum_{i'j'a'b'} G_{ii'} G_{jj'}
  \tau_{i'j'}^{a'b'} G^{-1}_{a'a} G^{-1}_{b'b} \\
  \lambda_a^i &\larrow \sum_{i'a'} G_{aa'} \lambda_{a'}^{i'} G^{-1}_{i'i} \\
  \lambda_{ab}^{ij} &\larrow \sum_{i'j'a'b'} G_{aa'} G_{bb'}
  \tau_{a'b'}^{i'j'} G^{-1}_{i'i} G^{-1}_{j'j}
\end{align}
\end{subequations}
with corresponding expressions for higher order excitations. This
completes the definition of the action $G\circ z$.

As a consequence of the transformation, $T\larrow T$,
$\Lambda\larrow\Lambda$, while $\ket{\phi}\larrow g\ket{\phi}$ and
$\bra{\tphi}\larrow\bra{\tphi} g^{-1}$, where
$g=\det(G_{ij})$. Clearly, $E[z]=E[G\circ z]$. As for the time-dependent functional
$S[z(\cdot)]$ it remains to check the part containing the time-derivative,
i.e.,
\begin{equation*}
  \int_0^T \bra{\tPsi}\pdiff{}{t}\ket{\Psi} \;\rmd t \longrightarrow \int_0^T
  \bra{\tPsi} g^{-1} \pdiff{}{t} g \ket{\Psi} \; \rmd t= \int_0^T
  \braket{\tPsi|\pdiff{}{t}|\Psi} + \pdiff{\ln g}{t} \; \rmd t,
\end{equation*}
so the integrand gains only a total time
derivative, 
\begin{equation*}
  S[z(\cdot)] \larrow S[G(\cdot)\circ z(\cdot)] = S[z(\cdot)] +
  \rmi\hbar \ln[g(T)/g(0)].
\end{equation*}
The last term is a constant with respect to variations vanishing at
the end points of $0\leq t \leq T$.

We now proceed to choose a gauge for the orbital rotation group.
Suppose we have a solution candidate $z_0(t)$, i.e., $\delta
S[z_0(\cdot)]=0$ for any variation of $z_0(t)$. For any choice
$G(t)\in \mathsf{G}$,
$z(t) = G(t)\circ z_0(t)$ is also a solution. To fix a unique solution
$z(t)$, we need to find a differential equation for $G(t)$ in terms of
$z_0(t)$ with a unique solution. To this end, consider the action of
$G(t)$ on the orbitals, i.e., $\Phi(t) = \Phi_0(t)G(t)$ and $\tPhi(t)
= G(t)^{-1}\tPhi_0(t)$. Writing $\eta_0(t) =
\tPhi_0(t)\dot{\Phi}_0(t)$ and $\eta(t) = \tPhi(t)\dot{\Phi}(t)$, we
have
\begin{equation*}
  G(t)\eta(t) =  \eta_0(t) G(t) + \dot{G}(t). 
\end{equation*}
We are free to choose the matrix elements $\dot{G}(t)_{ij}$ and
$\dot{G}(t)_{ab}$ of the nonzero blocks $G(t)_{\text{occ}}$ and
$G(t)_{\text{vir}}$, respectively, of $G(t)$ at will. Let $g(t)_{ij}$ and $g(t)_{ab}$ be
the matrix elements of some arbitrary matrices $g(t)_{\text{occ}}$ and $g(t)_{\text{vir}}$,
respectively, 
and require
\begin{align*}
  \dot{G}(t)_{\text{occ}} &= G(t)_{\text{occ}}
  g(t)_{\text{occ}} - \eta_0(t)_{\text{occ}}
  G(t)_{\text{occ}} \\
  \dot{G}(t)_{\text{vir}} &= G(t)_{\text{vir}} g(t)_{\text{vir}} - \eta_0(t)_{\text{vir}} G(t)_{\text{vir}}
\end{align*}
which defines $G(t)$ uniquely. Then,
\begin{equation*}
  \eta(t)_{\text{occ}} = g(t)_{\text{occ}}, \quad \text{and} \quad    \eta(t)_{\text{vir}} = g(t)_{\text{vir}}
\end{equation*}
so that any choice of $g(t)_\text{occ}$ and $g(t)_\text{vir}$, i.e.,
choice of gauge, implies a 
specific choice of transformation $G(t)$, and hence a solution
representant $z(t)$, all being equivalent \cite{Lubich2008}.

The simplest choice is probably $g(t)_\text{occ} \equiv 0$ and $g(t)_\text{vir} \equiv 0$, such that 
\begin{equation*}
  \eta(t) = \tPhi(t)\dot{\Phi}(t) = \begin{bmatrix} 0 &
    [\eta_{ia}] \\ [\eta_{ai}] & 0 \end{bmatrix}.
%  \label{eq:eta-structure}
\end{equation*}
To derive the differential equations corresponding to this gauge, one
needs to perform all possible variations $\delta z(t)$ adhering to the
constraint $(\tPhi \delta\Phi)_{\text{occ/vir}}=0$, i.e.,
$\braket{\tvphi_i|\delta{\vphi}_j} = \braket{\tvphi_a|\delta{\vphi}_b} =
0$. 

Note that $G(t)$ is simply a theoretical device that describes the
continuum of solutions $z(t)$. Using this device we derived conditions
on $\dot{z}(t)$ and the variations $\delta z(t)$ that picks exactly
\emph{one} of these solutions. $G(t)$ will not \emph{actually} be
solved for, since it has no physical value.

% Suppose we have found one solution $z(t) = (\lambda(t), \tau(t),
% \Phi(t), \tPhi(t))$ to the variational equations. Then $G(t)z(t)$ is a
% different but physically equivalent solution. We now show that this
% gauge degree of freedom can be eliminated by certain conditions on the
% allowed variations $\delta\Phi$ and $\delta\tPhi$. That is to say,
% using these conditions $G(t)$ is chosen in a unique way, thereby
% making the solution to the variational problems unique.

However, $\mathsf{G}$ as defined above is not the \emph{largest} group leaving
the functionals invariant. It turns out that the singles
amplitudes $\tau_i^a$ may be completely eliminated as well, corresponding to orbital
transformations on the form
\begin{equation*}
  G = e^{
  \begin{pmatrix}
    0 & 0 \\
    \tau_i^a & 0
  \end{pmatrix}}.
\end{equation*}
To see this, note that
\begin{equation*}
  \ket{\Psi} = e^{T}\ket{\phi} = e^{T-T_1} e^{T_1}\ket{\phi} =
e^{T-T_1}  \left(e^{T_1}c^\dag_1 e^{-T_1}\right)\cdots \left(e^{T_1}c^\dag_N
    e^{-T_1}\right) \ket{-},
\end{equation*}
and that the transformation $c^\dag_p \larrow \exp(T_1)c^\dag_p
\exp(-T_1)$ is equivalent to (see \ref{sec:technical})
\begin{equation*}
  \Phi \larrow \Phi e^{\begin{pmatrix} 0 &
      0 \\ \tau_i^a & 0\end{pmatrix}}, \quad  \tPhi \larrow e^{-\begin{pmatrix} 0 &
      0 \\ \tau_i^a & 0\end{pmatrix}} \tPhi. %\quad
%  \tphi \larrow \exp\left(\begin{bmatrix} 0 &
%      0 \\ -\tau_i^a & 0\end{bmatrix}\right) \tphi
\end{equation*}
The dual state $\bra{\tPsi}$ is invariant:
\begin{equation*}
  \bra{\tPsi} = \bra{\tphi}(1 + \Lambda)e^{-T} \larrow \bra{\tphi}(1 +
      e^{T_1}\Lambda e^{-T_1})e^{-T + T_1} = \bra{\tPsi},
\end{equation*}
where we have used that $\bra{\tphi}e^{T_1} = \bra{\tphi}$. The
bivariational functionals are then invariant under the described
transformation, and we may set $\tau_i^a\equiv 0$.

Observe, that we did \emph{not} transform the amplitudes
$\lambda_a^i$. It is tempting to assume that a transformation on the
form
\begin{equation*}
  G = e^{
  \begin{pmatrix}
    0 &  \sigma_a^i\\
    0 & 0
  \end{pmatrix}}
\end{equation*}
may achieve an elimination of $\lambda_a^i$. This is not the
case. There is no transformation
\begin{equation*}
  T = \sum_\mu \tau^\mu X_\mu \larrow T' = \sum_\mu (\tau')^\mu
  X_\mu
\end{equation*}
with the same truncation level that compensates for the transformation
\begin{equation*}
  \ket{\Psi} = e^T \ket{\phi} \larrow e^{S_1} e^{T'} e^{-S_1} e^{S_1}\ket{\phi}
  = e^{S_1} e^{T'} \ket{\phi},
\end{equation*}
i.e., $G$ induces higher-order excitations in $T$. The bivariational
functionals therefore cannot be made invariant under the
transformation, so $\lambda_a^i$ cannot be eliminated in this way.

However, it is easy to see, that if we \emph{do} include $\lambda_a^i$
as parameters the equations of motion will be overdetermined. The
presence of $T_1$ is compensated by the freely varying orbitals, but
the same is not true for $\Lambda_1$. We conclude that in the
orbital-adaptive CC, the dual state $\bra{\tPsi}$ would have
\emph{more} parameters than $\ket{\Psi}$. Correspondingly,
$\bra{\tPsi}$ and $\ket{\Psi}$ would not be one-to-one.

From now on, we therefore set \emph{both} $\tau_i^a$ and $\lambda_a^i$
identically equal to zero, in effect starting with a coupled-cluster
doubles, triples, etc, model with adaptive orbitals.

\subsection{Derivation of equations of motion}
\label{sec:eom-deriv}

Performing the variations with respect to the amplitudes is
straightforward (see \ref{sec:technical}), and leads to the
equations
\begin{subequations}
  \label{eq:amp-eq}
\begin{align}
  \rmi\hbar \dot{\tau}^\mu &=
  \pdiff{}{\lambda_\mu} \mathcal{E}_{H-\rmi\hbar
    D_0}[\lambda,\tau,\tPhi,\Phi]  =
  \braket{\tphi_\mu|e^{-T}(H-\rmi \hbar D_0)e^T|\phi} \\
  -\rmi\hbar \dot{\lambda}_\mu &=
  \pdiff{}{\tau^\mu} \mathcal{E}_{H-\rmi\hbar
    D_0}[\lambda,\tau,\tPhi,\Phi] =
  \braket{\phi|(1+\Lambda)e^{-T}[H - \rmi \hbar D_0,X_\mu]e^T|\phi}
\end{align}
\end{subequations}
which must hold for all $\mu\in\mathcal{I}$ included in the
approximation. Equations (\ref{eq:amp-eq}) are identical to Equations
(\ref{eq:response-eom}) except for the presence of $D_0$ due to the
changing orbitals.

Performing the variations with respect to the orbitals, we observe
that an arbitrary variation of the one-body function $\ket{\vphi_p}=\ket{\vphi_p(t)}$
(where we use ket notation for clarity) can be written
\begin{equation*}
  \delta \ket{\vphi_p} = P  \delta \ket{\vphi_p} + Q   \delta \ket{\vphi_p},
\end{equation*}
with $P = \Phi\tPhi=\sum_q \ket{\vphi_q}\bra{\tvphi_q}$ and $Q = 1-P$. Due to
the gauge choice $\eta^i_{j}=\eta^a_{b}=0$, we have for the occupied
and virtual orbitals
\begin{align*}
  \delta \ket{\vphi_i} &= \sum_b \epsilon^b_{i} \ket{\vphi_b} + Q
  \delta \ket{\vphi_i}\\\intertext{and} \delta \ket{\vphi_a} &= \sum_j
  \epsilon^a_{j} \ket{\vphi_j} + Q \delta \ket{\vphi_a},
\end{align*}
respectively, where $\epsilon^a_{i}$ are arbitrary independent
functions of time, and $Q\delta\ket{\vphi_i}$ is completely arbitrary
and independent from the $\epsilon^a_{i}$. We may therefore perform the
variations in two stages: First, we set $\delta\ket{\vphi_i} =
\epsilon(t)\ket{\vphi_a}$ for each $a,i$ separately (then exchange $i$
and $a$), and then finally we
set $\delta\ket{\vphi_p} = \ket{\theta} = Q\ket{\theta}$.

Beginning with the $P$-part of the variations, we note that the
variations in $\bra{\tvphi_p}$ are linked to those of $\ket{\vphi_p}$ due to the biorthogonality
constraint. We get
\begin{equation*}
  0 = \delta\braket{\tvphi_a|\vphi_i} =
  \braket{\delta\tvphi_a|\vphi_i} + \epsilon(t) \quad \Leftrightarrow \quad
  \delta\bra{\tvphi_a} = -\epsilon(t)\bra{\tvphi_i}.
\end{equation*}
It is useful to consider an arbitrary $\ket{u}$ and $\bra{v}$, whose
variations are
\begin{equation*}
  \delta\ket{u} = \epsilon c^\dag_a \tilde{c}_i \ket{u} \quad\text{and}\quad
  \delta\bra{v} = -\epsilon \bra{v} c^\dag_a \tilde{c}_i,
\end{equation*}
so that we have the general equation
\begin{equation*}
  \delta\braket{u|A|v} = \epsilon \braket{u|[A,c^\dag_a \tilde{c}_i]|v} + \braket{u|(\delta A)|v},
%  \label{eq:commutator-delta}
\end{equation*}
where $A$ is any operator that may depend explicitly on the
orbitals. (If $A$ does not depend on the orbitals, then $\delta A=0$.)
For example, the Hamiltonian does not depend explicitly on the
orbitals, while
\begin{equation*}
  \delta D_0 = \delta \sum_q \dot{c}^\dag_q \tilde{c}_q =
  \left(\dot{\epsilon}c^\dag_a + \epsilon \dot{c}^\dag_a\right)\tilde{c}_i
    -\epsilon \dot{c}^\dag_a \tilde{c}_i = \dot{\epsilon}c^\dag_a \tilde{c}_i.
\end{equation*}
We also note that 
\begin{equation*}
  \delta \sum_\mu \dot{\tau}^\mu \braket{\tPsi| X_\mu |\Psi} = 0,
\end{equation*}
since the expectation value of $X_\mu$ only depends on the amplitudes
and not the orbitals. We compute the variation in $\mathcal{S}$:
\begin{align}
  \delta\mathcal{S} &= \delta \int_0^T \braket{\tPsi|\rmi\hbar D_0 +
    \rmi\hbar \sum_\mu \dot{\tau}^\mu X_\mu - H|\Psi} \; \rmd t \notag\\
                    &= \int_0^T \epsilon(t) \braket{\tPsi|[\rmi\hbar D_0
    -
    H, c^\dag_a \tilde{c}_i]|\Psi} + \rmi\hbar
  \dot{\epsilon}(t)\braket{\tPsi|c^\dag_a\tilde{c}_i|\Psi} \; \rmd t
\label{eq:delta-s}  \\
  &= \int_0^T \epsilon(t)\left( \braket{\tPsi|[\rmi\hbar D_0 -
    H, c^\dag_a \tilde{c}_i]|\Psi} - \rmi\hbar \dot{\rho}^i_{a} \right)
\; \rmd t. \notag
\end{align}
Requiring $\delta\mathcal{S}=0$ for all $\epsilon(t)$ implies that the
integrand must vanish. Using the gauge condition, $D_0$ becomes 
\begin{equation*}
  D_0 = \sum_{jb} \eta^j_{b} c^\dag_j \tilde{c}_b + \sum_{jb} \eta^b_{j}
  c^\dag_b \tilde{c}_j,
\end{equation*}
and only the first sum survives in the commutator in
Eqn.~(\ref{eq:delta-s}). We obtain the equation
\begin{equation}
  \rmi\hbar \sum_{bj}  \braket{\tPsi|[c^\dag_j \tilde{c}_b, c^\dag_a
    \tilde{c}_i]|\Psi}\eta^b_{j} = \rmi\dot{\rho}^i_a +
  \braket{\tPsi|[H, c^\dag_a\tilde{c}_i]|\Psi},
\label{eq:pspace-almost-there}
\end{equation}
which  is a linear equation for $\eta^b_{j} =
\braket{\tvphi_b|\dot{\vphi}_j}$. It is readily verified that
\begin{equation*}
  \rho^i_{a} = \braket{\tPsi|c^\dag_a\tilde{c}_i|\Psi} =
  \braket{\tphi|(1 + \Lambda)e^{-T} X_i^a e^T|\phi} = \lambda_a^i
  \equiv 0,
\end{equation*}
and that
\begin{subequations}
  \label{eq:commutators}
  \begin{align}
    [c^\dag_p \tc_q, c^\dag_a \tc_i] &= \delta^q_a c^\dag_p \tc_i -
    \delta^i_{p} c^\dag_a \tc_q \\
    [c^\dag_p c^\dag_r \tc_s \tc_q, c^\dag_a \tc_i] &= \delta^q_{a}c^\dag_p
    c^\dag_r \tc_s \tc_i - \delta^s_{a} c^\dag_p c^\dag_r \tc_q \tc_i +
    \delta^i_{r} c^\dag_a c^\dag_p \tc_s \tc_q - \delta^i_{p} c^\dag_a
    c^\dag_r \tc_s \tc_q.
  \end{align}
\end{subequations}
The coefficient matrix of Eqn.~(\ref{eq:pspace-almost-there}) becomes
\begin{equation}
  A^{ib}_{aj} \equiv \braket{\tPsi|[c^\dag_j \tilde{c}_b, c^\dag_a
    \tilde{c}_i]|\Psi} = \delta^b_{a}\rho^i_{j} - \delta^i_{j}\rho^b_{a}.
  \label{eq:A-def}
\end{equation}
In total, we get a linear equation for $\eta^b_{j}$ that reads
\begin{equation*}
  \rmi\hbar \sum_{bj} A^{ib}_{aj}\eta^b_{j} = \sum_p \rho^i_{p} h^p_{a} - \sum_q
  \rho^q_{a} h^i_{q} + \frac{1}{2}\left[ \sum_{prs} \rho^{is}_{pr}
    u^{pr}_{as} - \sum_{rqs} \rho^{qs}_{ar} u^{ir}_{qs}\right],
\end{equation*}
where we used Eqns.~(\ref{eq:commutators}) and the anti-symmetry 
$u^{pr}_{qs} = -u^{pr}_{sq}$.

We now turn to the variation $\delta\ket{\vphi_a} =
\epsilon(t)\ket{\vphi_i}$, which implies $\delta\bra{\tvphi_i} =
-\epsilon(t)\bra{\tvphi_a}$. The calculation is completely analogous
to the previous case, so we simply state the result,
\begin{equation*}
  - \rmi\hbar \sum_{bj} A^{ja}_{bi}\eta^j_b = \sum_p \rho^a_{p} h^p_{i} - \sum_q
  \rho^q_{i} h^a_{q} + \frac{1}{2}\left[ \sum_{prs} \rho^{as}_{pr}
    u^{pr}_{is} - \sum_{rqs} \rho^{qs}_{ir} u^{ar}_{qs}\right] + \rmi\hbar\dot{\rho}^a_{i}.
\end{equation*}
Unlike $\rho_a^i$, the coefficients $\rho^a_i$ do not vanish
identically, except for in the doubles only approximation to be
considered in Section~\ref{eq:doubles}.

Having derived the equations of motion for the $P$-part of
$\dot{\vphi}_p$ (and therefore also $\dot{\tvphi}_p$), we now turn to
the $Q$-part. We perform an arbitrary variation
$\delta\bra{\tvphi_{p'}} \equiv \bra{\theta} = \bra{\theta}Q$, i.e.,
$\braket{\theta|\vphi_q}=0$ for all $q$. These variations are therefore
independent from the corresponding variations in $\ket{\vphi_q}$. 

It is convenient to use $\mathcal{S}$ on the form (\ref{eq:the-functional-2}) to derive the
variational equations. Recall that the reduced density matrix elements
$\rho^q_p$ and $\rho^{qs}_{pr}$ are independent of the orbitals, since
they are computed only using the anti-commutator
(\ref{eq:bi-anti-comm}). Thus, the only quantities that vary are the
one- and two-body matrix elements $h^p_q$, $\eta^p_q$ and $u_{qs}^{pr}$:
\begin{align}
  \delta\mathcal{S} &= \delta\int_0^T \sum_{pq}\rho^q_p (\rmi\hbar
  \eta^p_q - h^p_q) - \frac{1}{4}\sum_{prqs} \rho^{qs}_{pr} u^{pr}_{qs} \;
  \rmd t \notag \\
  &= \int_0^T \sum_q \rho^q_{p'} \braket{\theta|(\rmi\hbar\pdiff{}{t} -
    h)|\vphi_q} - \frac{4}{4}\sum_{qrs}\rho^{qs}_{p'r}
  \braket{\theta\tvphi_r|u|\vphi_q\vphi_s} \; \rmd t   \label{eq:varvar}
\\
  &= \int_0^T \bra{\theta}\left[ \sum_{q}
    \rho^q_{p'}(\rmi\hbar\ket{\dot{\vphi_q}}-h\ket{\vphi_q}) -
    \sum_{qrs} \rho^{qs}_{p'r} \braket{\;\cdot\;\tvphi_r|u|\vphi_q\vphi_s}
  \right] \; \rmd t \notag
\end{align}
From line 1 to line 2 we used the symmetries $u^{pr}_{qs} = u^{rp}_{sq}$ and
$\rho^{qs}_{pr} = \rho^{sq}_{rp}$, and the two-electron integrals in the
last two lines are \emph{not} anti-symmetrized. We define mean-field
potentials $W^r_s$ by
\begin{equation*}
\braket{\;\cdot\;\tvphi_r|u|\vphi_q\vphi_s} \equiv \int \tvphi_r(x')
u(x,x') \vphi_q(x) \vphi_s(x') \; \rmd x \equiv W^r_s \ket{\vphi_q}.
%\label{eq:mean-field}
\end{equation*}
Since Eqn.~(\ref{eq:varvar}) must hold for \emph{all} $\bra{\theta} =
\bra{\theta}Q$, we get the equation
\begin{equation*}
  \rmi\hbar \sum_q \rho^q_p Q \pdiff{}{t}\ket{\vphi_q} = \sum_q
  \rho^q_p Q h\ket{\vphi_q} + \sum_{qrs} \rho^{qs}_{pr} Q W^r_{s}
  \ket{\vphi_q} \quad \text{for all $p$}.
%  \label{eq:q-space-ket}
\end{equation*}
Performing the variation $\delta\ket{\vphi_q} = \ket{\theta} =
Q\ket{\theta}$ is completely analogous, and gives
\begin{equation*}
  -\rmi\hbar \sum_p \rho^q_{p} \left(\pdiff{}{t}\bra{\tvphi_p}\right)Q = \sum_p
  \rho^q_{p} \ket{\tvphi_p} h Q + \sum_{prs} \rho^{qs}_{pr}
  \bra{\tvphi_p}W^r_{s} Q \quad\text{for all $q$},
%  \label{eq:q-space-bra}
\end{equation*}
where the minus sign comes from integration by parts.

\subsection{The doubles approximation: OATDCCD}
\label{eq:doubles}

The simplest non-trivial OATDCC case is the doubles approximation
(OATDCCD). The wavefunction parameters are, in addition to the
orbitals $\tPhi$ and $\Phi$, the amplitudes $\tau=(\tau_{ij}^{ab})$ and
$\lambda=(\lambda_{ab}^{ij})$. In \ref{sec:algebraic}, a complete
listing of the algebraic expressions needed to evaluate the equations
of motion is given. In particular, the only nonzero elements of the
one-body reduced density matrix are $\rho^j_i$ and $\rho^b_a$, which
simplifies the equations of motion. The amplitude equations read
\begin{subequations}
\label{eq:eom}
\begin{align}
  \rmi\hbar \dot{\tau}_{ij}^{ab} &= \pdiff{}{\lambda_{ab}^{ij}}
  \mathcal{E}_H[\lambda,\tau,\tPhi,\Phi] = \braket{\tphi_{ij}^{ab}|e^{-T} H e^T|\phi}\label{eq:eom-1} \\
  -\rmi\hbar \dot{\lambda}_{ab}^{ij} &= \pdiff{}{\tau^{ab}_{ij}}
  \mathcal{E}_H[\lambda,\tau,\tPhi,\Phi] =
  \braket{\tphi|(1+\Lambda)e^{-T} [H,X^{ij}_{ab}]
    e^T|\phi}\label{eq:eom-2}. \\ \intertext{The $P$-space orbital
    equations read}
%   \rmi\hbar \sum_{bj} A_{ia,bj}\eta_{bj} &= \sum_p
%   \rho_{pi} h_{pa} - \sum_q \rho_{aq} h_{iq} + \frac{1}{2}\left[
%     \sum_{prs} \rho_{pris} u_{pras} - \sum_{rqs} \rho_{arqs}
%     u_{irqs}\right],
%   \label{eq:eom-3} \\
%   - \rmi\hbar \sum_{bj} A_{bj,ia}\eta_{jb} &= \sum_p \rho_{pa} h_{pi}
%   - \sum_q \rho_{iq} h_{aq} + \frac{1}{2}\left[ \sum_{prs} \rho_{pras}
%     u_{pris} - \sum_{rqs} \rho_{irqs} u_{arqs}\right] +
%   \rmi\hbar\dot{\rho}_{ia}.
  \rmi\hbar \sum_{bj} A^{ib}_{aj}\eta^b_{j} &= \sum_j \rho^i_{j} h^j_{a} - \sum_b
  \rho^b_{a} h^i_{b} + \frac{1}{2}\left[ \sum_{prs} \rho^{is}_{pr}
    u^{pr}_{as} - \sum_{rqs} \rho^{qs}_{ar}
    u^{ir}_{qs}\right] \label{eq:eom-3} \\
  - \rmi\hbar \sum_{bj} A^{ja}_{bi}\eta^j_b &= \sum_b \rho^a_{b} h^b_{i} - \sum_j
  \rho^j_{i} h^a_{j} + \frac{1}{2}\left[ \sum_{prs} \rho^{as}_{pr}
    u^{pr}_{is} - \sum_{rqs} \rho^{qs}_{ir} u^{ar}_{qs}\right] .
\label{eq:eom-4} \\ \intertext{Finally, the $Q$-space orbital
  equations are}
%  \rmi\hbar \sum_q \rho_{pq} Q \pdiff{}{t}\ket{\vphi_q} &= \sum_q
%  \rho_{pq} Q h\ket{\vphi_q} + \sum_{qrs} \rho_{prqs} Q W_{rs}
%  \ket{\vphi_q} \label{eq:eom-5} \\
  \rmi\hbar \sum_q \rho^q_p Q \pdiff{}{t}\ket{\vphi_q} &= \sum_q
  \rho^q_p Q h\ket{\vphi_q} + \sum_{qrs} \rho^{qs}_{pr} Q W^r_{s}
  \ket{\vphi_q} \quad  \label{eq:eom-5} \\ 
  -\rmi\hbar \sum_p \rho^q_{p} \left(\pdiff{}{t}\bra{\tvphi_p}\right)Q &= \sum_p
  \rho^q_{p} \ket{\tvphi_p} h Q + \sum_{prs} \rho^{qs}_{pr}
  \bra{\tvphi_p}W^r_{s} Q .
%  -\rmi\hbar \sum_p \rho_{pq} (\pdiff{}{t}\bra{\tvphi_p})Q &= \sum_p
%  \rho_{pq} \ket{\tvphi_p} h Q + \sum_{prs} \rho_{prqs}
%  \bra{\tvphi_p}W_{rs} Q.
  \label{eq:eom-6}
\end{align}
\end{subequations}
The coefficients $A^{ib}_{aj}$ are defined in
Eqn.~(\ref{eq:A-def}). The right-hand sides of
Equations~(\ref{eq:eom-1}) and (\ref{eq:eom-2}), which are polynomials
in terms of $\tau_{ij}^{ab}$, $\lambda_{ab}^{ij}$, $h^p_q$ and
$u^{pr}_{qs}$, can be found in the Appendix. Note that the right-hand
sides of the equations are \emph{identical} to the one used in
standard CCD calculations for the ground state energy, since the
operator $D_0$ is eliminated due to $\rho^a_i=\rho^i_a=0$. Existing computer
codes may be helpful for implementations.

Since $D_0$ drops from (\ref{eq:eom-1}) and
(\ref{eq:eom-2}), the right hand sides can be evaluated independently
of Eqns.~(\ref{eq:eom-3}) to (\ref{eq:eom-6}). Note that to evaluate
$\dot{\tPhi}$ and $\dot{\Phi}$, $\eta$ must be solved for in addition
to $Q\ket{\dot{\varphi}_q}$ and $\bra{\dot{\tvphi}_p} Q$. The results are
assembled according to 
\begin{align*}
  \ket{\dot{\vphi}_q} &= (P+Q)\ket{\dot{\vphi}_q} = \sum_p \ket{\vphi_p}\braket{\tvphi_p|\dot{\vphi}_q}
  + Q\ket{\dot{\vphi}_q} = \sum_p \eta^p_q \ket{\vphi_p} +
  Q\ket{\dot{\vphi}_q} \\
  \bra{\dot{\tvphi}_p} &= \bra{\dot{\tvphi}_p}(P+Q) = \sum_q \braket{\dot{\tvphi}_p|\vphi_q}\bra{\tvphi_q}
  + \bra{\dot{\tvphi}_q}Q = -\sum_q \eta^p_q \bra{\tvphi_q} +
  \bra{\dot{\tvphi}_p}Q 
\end{align*}

We note that the $Q$-space equation for $\ket{\vphi_p}$ is formally
\emph{identical} to the orbital equation of MCTDHF (see 
\ref{sec:mctdhf-derivation}), and the equation for 
$\bra{\dot{\tvphi}_p}$ is formally identical to the complex
conjugate. However, in the CC case the matrices $u_{pr}^{qs}$, $h^p_{q}$ and
$\rho^q_p$ are \emph{not exactly Hermitian}. Therefore, the two equations are
only complex conjugates of each other to within an approximation, and
both must be propagated.

We will now consider the computational cost of evaluating the time
derivatives in a computer implementation. We will in the following
assume a grid-based discretization of single-particle space using in
total $N_b$ points. In particular, integrals are evaluated as sums
with $N_b$ elements.

We consider first the computation of the amplitude equations
(\ref{eq:eom-1}) and (\ref{eq:eom-2}), assuming that $h^p_q$ etc are
available in computer memory. It is easy to see, from
Equations~(\ref{eq:lambda-derivatives}) and
(\ref{eq:tau-derivatives}), that by brute-force summation the
worst-scaling terms require $O(N^4(L-N)^4)=O(L^8)$ operations for
computing the totality of derivatives, which is a conservative
estimate since $N<L$. Existing CC codes typically reduce this to $O(L^6)$ by
clever use of intermediate variables \cite{Crawford2000}.

The $P$-space orbital equations (\ref{eq:eom-3}) and (\ref{eq:eom-4})
are linear equations where a vector of dimension $O(L^2)$ is to be
solved for. This requires at most $O(L^6)$ operations. The right-hand
side is dominated by the two-body terms, which cost $O(L^5)$ in total
to compute.

We next turn to the $Q$-space orbital equations (\ref{eq:eom-5}) and
(\ref{eq:eom-6}), which can be viewed as differential equations for
matrices of dimension $N_b\times L$ and $L\times N_b$,
respectively. The cost analysis is identical for the two. The matrix
$\rho^{q}_p$ needs to be inverted, a step of at most $O(L^3)$
cost. Multiplying Eqn.~(\ref{eq:eom-5}) by the inverse matrix elements
of $\rho^q_p$ shows that the $Q$-part of $\rmi\hbar\ket{\dot{\vphi}_p}$
can be computed as the sum of $Qh\ket{\vphi_p}$ and a two-body
mean-field term which clearly dominates the computation. The cost of
this term is $O(L^3 N_b)$ plus $O(L^2 N_b)$ for the multiplication of
the result with $\rho^{-1}$.

Unlike standard CC calculations, the one- and two-electron integrals
$h^p_q$ and $u^{pr}_{qs}$ must be updated \emph{at each time
  $t$}. This is similar to the situation in MCTDHF theory. Moreover,
equations (\ref{eq:eom-3})--(\ref{eq:eom-6}) are formulated in terms
of the reduced one- and two-electron matrices which need to be
computed.

The matrices $\rho^{q}_p$ and $\rho^{qr}_{ps}$ cost less than the
evaluation of the orbital equations right-hand sides in
total, and $h^p_q$ is relatively cheap to compute. However, the
two-electron integrals and the mean-field functions $W^r_s$ are
costly. The computation of all the mean-fields, which are local
functions, costs $O(L^2N_b^2)$. Since
$u^{pq}_{qs}=\braket{\tvphi_p|W^{r}_s|\vphi_q}$, the computation of the
two-electron integrals costs an additional $O(L^4 N_b)$ operations.

The computation of $W^r_s$ is in fact very expensive, being similar 
in cost to computing two-particle integrals, i.e., six-dimensional
integrals in realistic calculations. This problem is ubiquitous for
all time-dependent mean-field calculations, and a common approach is
to employ some low-rank expansion for the interaction potential
$u(x,x')$, which needs to be sampled at the grid points $\xi_k$,
$k=1,\cdots,N_b$. The resulting matrix $v(\xi_k,\xi_{k'})$ is symmetric, with
eigenvalue decomposition
\begin{equation}
  v(\xi_k,\xi_{k'}) = \sum_{m=1}^{N_b} \lambda_m f_{m}(\xi_k) f_{m}(\xi_{k'}),
  \label{eq:lorank}
\end{equation}
where $f_m$ is the eigenvector belonging to $\lambda_m$, the latter
arranged in decreasing order. The optimal (in the 2-norm) $M$-term
approximation to $v(\xi_k,\xi_{k'})$ is then obtained by truncating
Eqn.~(\ref{eq:lorank}) after $M$ terms. Oftentimes, only a small
number $M\ll N_b$ terms are needed. Moreover, if $N_b$ is increased,
$M$ may typically be held fixed. The mean-fields then become
\begin{equation*}
  W^r_s(\xi_k) \approx \sum_{m=1}^M \lambda_m v_{m}(\xi_k) \braket{\tvphi_r|v_m|\vphi_s},
\end{equation*}
which reduces the cost of computing $W_s^r$ to $O(N_b L^2)$ for the
inner products plus $O(L^2 N_b M)$ for the summation, reducing the
cost proportionally to the fraction $M/N_b$ of modes included in
Eqn.~(\ref{eq:lorank}).

To sum up, we see that the cost of evaluating the right-hand sides of
the equations of motion is dominated by the computation of the $W^r_s$
and $u^{qr}_{ps}$ and of the evaluation of the amplitude equations,
costing $O(N^4(L-N)^4)$ (if no optimization is done), and
being the only terms that increase in complexity with the number $N$
of particles. This should be contrasted to MCTDHF calculations, where
the number of amplitudes grow \emph{exponentially} with $N$.

\section{Further properties of OATDCC}

\subsection{Relations to other methods}
\label{sec:method-relations}

It is instructive to consider special cases of the OATDCC method and
relate these to other, well-known wavefunction approximations.

In the case where \emph{all} excitation levels are included, we have
seen that the CC ansatz becomes the FCI ansatz within the chosen
basis. Therefore, the MCTDHF and OATDCC methods are equivalent in this
limit. In \ref{sec:mctdhf-derivation} an explicit derivation of MCTDHF
using the bivariational principle is carried out. Note, however, that
the gauge conditions are different, i.e., the orbitals produced are
not identical. This stems from the fact that the CC wavefunction is
normalized according to $\braket{\tphi|\Psi} = 1$ at all
times. However, the orbitals in the two methods span the same space,
and the wavefunctions are identical except for normalization. At each
time $t$, the OATDCC and MCTDHF wavefunctions produce the \emph{same}
expectation value functional for any observable.

At the other end of the hierarchy, we find the trivial case where
there are no amplitudes at all, i.e., we take $L=N$ and $\Lambda = T =
0$. In that case, the OATDCC energy expectation functional becomes
\begin{align*}
  \mathcal{E}_H[\Phi,\tPhi] &= 
   \braket{\tphi|H|\phi} =
  \braket{\tphi|\left( \sum_{pq} h^p_q c^\dag_p \tc_q + \frac{1}{4}\sum_{pqrs}
    u^{pr}_{qs} c^\dag_p c^\dag_r \tc_s \tc_q\right)|\phi} 
\\
&= \sum_{p}\braket{\tvphi_p|h|\vphi_p} + \frac{1}{2} \sum_{pr}
\braket{\tvphi_p\tvphi_r|u|\vphi_p\vphi_r}_{AS}.
\end{align*}
This is the Hartree--Fock energy functional (when $\tPhi = \Phi^\mathsf{H}$). Thus,
the conditions for $\delta \mathcal{E}_H = 0$ are the Hartree--Fock
equation and its complex conjugate. The action functional becomes
 \begin{align*}
   \mathcal{S}[\Phi,\tPhi] &= 
   \int_0^T \braket{\tphi|(\rmi\hbar D_0 - H)|\phi} \; \rmd t \\ 
 &= \int_0^T \rmi\hbar \sum_p \braket{\tvphi_p|\dot{\vphi}_p} -
 \sum_{p}\braket{\tvphi_p|h|\vphi_p} - \frac{1}{2} \sum_{pr}
 \braket{\tvphi_p\tvphi_r|u|\vphi_p\vphi_r}_{AS} \;
 \rmd t.
 \end{align*}
This is the time-dependent Hartree--Fock (TDHF) functional. Computing
the variation with respect to $\tvphi_q$ we get the TDHF equations of
motion, and the variation with respect to $\vphi_q$ gives the complex
conjugate, showing that the $T=\Lambda=0$ case is indeed equivalent to
TDHF. 

We also note that the OATDCCD approximation is equivalent to MCTDHF
whenever $N=2$. Moreover, some combinations of $L$ and $N$ also give
equivalence, for example $L=N+2$ since there are no triple excitations
defined. 

Finally, in the absence of interactions, the Hamiltonian is a pure
one-body Hamiltonian. One can easily show that the choice $\rmi\hbar \dot{\Phi}
= H\Phi$, $-\rmi\hbar\dot{\tPhi} = \tPhi H$ and $\dot{\lambda}_\mu =
\dot{\tau}^\mu = 0$ gives $\mathcal{S}[\lambda,\tau,\tPhi,\Phi] = 0$
and a stationary $\mathcal{S}$. This is the
\emph{exact} solution to the dynamics for any initial condition. (The
gauge conditions in this case are chosen differently from earlier: $\eta^i_j =
\braket{\tvphi_i|\dot{\vphi}_j} = \braket{\tvphi_i|H^{(1)}|\vphi_j}/\rmi\hbar$
and $\eta^a_b = \braket{\tvphi_a|H^{(1)}|\vphi_b}/\rmi\hbar$.)

In OATDCC the evolution of the orbitals are chosen to
variationally optimize the action functional. As early as 1978,
Hoodbhoy and Negele \cite{Hoodbhoy1978} discussed a time-dependent CC
approach using an \emph{explicit} dependence of time in the
orthonormal single-particle functions. This would correspond to using
the following functional to define the evolution:
\begin{equation*}
  \mathcal{S}_\text{H--N}[\tau,\lambda] = \int_0^T \rmi\hbar
  \lambda_\mu \dot{\tau}^\mu -
  \braket{\tphi|(1+\Lambda)e^{-T}(H-\rmi\hbar D_0(t))e^T|\phi}\; \rmd t.
\end{equation*}
The $D_0$ operator simply is a correction in the standard CC
Lagrangian due to a moving basis.  Hoodbhoy and Negele suggested
computing the time-dependence of $\Phi$ using, say, TDHF, i.e., a TDHF
calculation is first performed, and the output is fed into
$\mathcal{S}_\text{H--N}$. However, this approach would have inferior
approximation properties compared to OATDCC while at the same time
being only marginally easier to evolve in time. To see this, consider
the fact that the TDHF solution would depend \emph{only} on the
initial choice of the orbitals, and \emph{not} on the full state at
time $t$ as in the OATDCC approach. The OATDCC orbitals will
generally differ substantially from the TDHF solution, since their
motion is computed from the wavefunctions \emph{at time $t$}. The
Hoodbhoy--Negele TDHF approach neglects the correlation effects built into
the wavefunction during the evolution. As for the computational cost,
note that TDHF needs the computation of the two-particle integrals, so
one gains very little, if nothing, at simplifying to TDHF for the
orbitals.

\subsection{Approximation properties}

We now ask: what kinds of systems can we expect to be able to treat
with OATDCC, and what systems cannot be expected to give good results?

The usual CC ansatz (with fixed orbitals) is based on a single reference determinant,
incorporating correlations through the cluster operator. For good
results, the reference determinant should be a ``large'' part of the
wavefunction. In the language of computational chemistry, dynamic
correlation (which by definition is due to the interparticle
interactions) must be dominating, while static correlation (arising
from degeneracies in the spectrum) should be
small.

For a \emph{dynamical} calculation we must correspondingly require
that, for all $t\geq 0$, the wavefunction is a single-reference type
state. This is reasonable whenever the \emph{initial condition}
is of such type: intuitively, the Hamiltonian cannot generate static correlation
since the only source of correlation from dynamics is the
interparticle interaction. This is actually observed in the numerical
experiment in Section~\ref{sec:experiment}. 

However, standard CC is known to perform adequately even in the
presence of static correlation, which gives reason to believe that the
same holds true for OATDCC calculations. In any case, OATDCCD should
be much better than a singles-and doubles truncation of MCTDHF due to
size-consistency, even though the latter is a ``true''
multiconfigurational method, where all the basis determinants are
independent.

It is worthwhile to note, that for systems with spin, the OATDCC
ansatz is not an eigenfunction of the total spin; only of the total
spin projection along some preselected direction in space $z$. This
is, however, not a problem in general if $H$ commutes with total spin:
all expectation values for spin-independent observables are the same
as for the properly ``spin-symmetrized'' wavefunction.

\subsection{Non-feasibility of imaginary time relaxation}

We have not yet discussed choices of initial conditions for OATDCC
calculations. Typically, one would like to start in the ground state
of the system under consideration, or a state closely related to
this. Indeed, the orbital-adaptive CC ansatz could in principle be
used for ground-state calculations in the first place, just like
MCTDHF actually is a time-dependent version of the multi-configuration
Hartree--Fock (MCHF) for computing eigenvalues of $H$.

The ground state is a critical point for $\mathcal{E}_H$, and in
MCTDHF theory this can be computed using imaginary
time propagation, that is to say, formally replacing the time $t$ with
$-\rmi s$; so-called Wick rotation. Asymptotically, as $s\rightarrow \infty$, a
critical point of the variational energy is obtained. This is a quite
robust procedure for variational approximations. For non-variational
methods like coupled-cluster, the situation is, literally,
more complex. 

To see this, consider again local coordinates
$z(t)\in\mathbb{C}^n$. The equations of motion for $z(t)$ are
\emph{analytic} in both $z$ and $t$. Thus, the energy is an
\emph{analytic} function of $t$, which is conserved,
$\rmd\mathcal{E}[z(t)]/\rmd t \equiv 0$. Thus, the energy is a
constant analytic function, even for complex $t$. Thus, the energy
will not decay exponentially when the system is propagated in
imaginary time.

We conclude that computing the ground state using imaginary time
propagation is not straightforward and requires separate
study. Instead, quasi-Newton schemes like those already used for
standard CC could be used, but we will not investigate this further in
the present article.

Suppose the ground state is desired as initial condition. One option
is to perform a Hartree--Fock calculation to generate a set of
orbitals, and then perform a CCD calculation within this basis. This is
the standard practice for molecular calculations, and even though not
an \emph{exact} critical point of the OATDCC energy, it should be a
suitable starting point for dynamics calculations. In our numerical
experiment in Section \ref{sec:experiment} we choose a similar
approach.

\section{A numerical experiment}
\label{sec:experiment}

\subsection{Outline and model system}
\label{sec:experiment-outline}

A numerical experiment on a model system mimicking electron-atom
collision has been performed in order to test the OATDCCD method
against the standard MCTDHF method. (Recall that OATDCCD is an
approximation to MCTDHF.) Generic MCTDHF and OATDCCD codes has been
written from scratch. The MCTDHF code was tested against numerical
experiments reported in the literature \cite{Zanghellini2004} and
found to agree perfectly with these. The OATDCCD code uses the
algebraic expressions listed in \ref{sec:algebraic} for the
equations of motion. The code is tested against MCTDHF computations
for special combinations of $L$ and $N$ where the two ansätze are
equivalent. Perfect agreement was found, indicating the correctness of
the implementation.

The numerical experiment consists of two phases: (1) preparation of
the initial wavefunction, and (2) propagation of this state from $t=0$
to $t=t_\text{final}$ using both MCTDHF and OATDCC while monitoring
some observables. In particular the energy $\mathcal{E}_H$ should be
conserved at all times.

The test system is defined as follows. Consider a model consisting of
$N$ electrons in one spatial dimension. The orbitals are 
functions $\varphi_p(x,s)$, where $x\in\mathbb{R}$ is the spatial
position and $s\in\{-\frac{1}{2}, +\frac{1}{2}\}$ is the quantum
number of the projection of the electron spin along some arbitrary
axis.

The particles interact via a smoothed Coulomb force (for simplicity)
and an external Gaussian well potential. The Hamiltonian of the
system has one-body part
\begin{equation*}
  h = -\frac{1}{2} \pdiff{^2}{x^2} + V(x), \quad V(x) = -V_0 e^{-x^2/2a^2},
\end{equation*}
where $V_0 = 7$ and $a = 1.5$ are the parameters of the Gaussian
well. The smoothed Coulomb interaction is given by
\begin{equation*}
  u(x_1, x_2) = \frac{\lambda}{\sqrt{|x_1-x_2| + \delta^2}},
\end{equation*}
and we use parameters $\lambda = 1$ and $\delta = 0.2$. These are
reasonable parameters for, say, a quantum wire model \cite{Reimann2002}.

The complete Hamiltonian is seen to commute with any spin operator. We
remark that if (1) the initial orbitals are spin-orbitals on the form
$\varphi_p(x,s) = \psi_p(x)\chi_{\sigma_p}(s)$ where $\chi_\sigma$ is
a spinor basis function, and if (2) the initial wavefunction is an
eigenfunction for the \emph{total} spin projection operator, the equations
of motion (\ref{eq:eom}) preserve these properties. That
is to say, under these conditions the orbitals are always on product
form
\begin{equation*}
  \varphi_p(x,s,t) = \psi_p(x,s,t)\chi_{\sigma_p}(s),
\end{equation*}
and the wavefunction remains an eigenfunction of the total spin
projection.

\subsection{Discretization and propagation scheme}
\label{sec:experiment-discr}

We discretize the one-particle coordinates by introducing a standard
discrete Fourier transform-based discretization over the interval
$[-R,R]$ using $N_\text{grid} = 64$ points \cite{Tal-Ezer1984}. The
total number of basis functions is then $N_b = 2N_\text{grid} =
128$. The kinetic energy operator is evaluated using the fast Fourier
transform (FFT) and is highly efficient and accurate. In our
calculations, we set $R=15$.  The
orbital matrices $\tPhi$ and $\Phi$ become standard matrices of
dimension $L\times N_b$ and $N_b\times L$, respectively, giving a
simple representation in the computer code.

For propagation, we choose a variational splitting scheme
\cite{Koch2011}. Variational splitting is a generalization of the
standard split-step scheme used for brute-force grid discretizations
of few-body problems \cite{Tal-Ezer1984}. Using this scheme, the time
step $\Delta t$ can be chosen relatively large and independently of
the grid spacing $\Delta x$. Variational splitting is most easily
described in terms of a \emph{time-dependent} Hamiltonian
$\tilde{H}(t) = T + \sum_{n=-\infty}^\infty \delta(t - (n+1/2)\Delta
t) (H - T)$, where $T$ is the kinetic energy operator and $H-T$ is the
remaining potential terms. A time step then consists of three steps:
(1) Propagation using kinetic energy \emph{only} as a time step
$\Delta t/2$, (2) Propagation using $H-T$ \emph{only} a time step
$\Delta t$ using a fourth order Runge--Kutta for simplicity, and
finally step (1) is repeated. The key point is that the stability of
the scheme becomes insensitive to $\Delta x$, and that step (1) is
evaluated exactly without involving the amplitudes at all. The local
error is of order $\Delta t^3$. A simple integration of
Eqns.~(\ref{eq:eom}) using, say, Runge--Kutta will require a time step
$\Delta t \sim \Delta x^2$. In multi-configurational time-dependent
Hartree calculations, other ways of eliminating this stability problem
than variational splitting is often used, e.g., the constant
mean-field scheme \cite{Beck2000}. However, this requires more coding
effort than variational splitting which is sufficient for our modest
purposes.

\subsection{Preparation of initial wavefunction}

We need initial wavefunctions for both the MCTDHF and OATDCCD
ansätze. The former is computed as follows.

The parameters of the Gaussian well potential and interaction
potential are experimentally chosen so that they support an $N=4$
ground state $\ket{\Psi_4}$ with total spin projection zero. This
ground state is computed in the MCTDHF scheme using imaginary time
propagation of the equations of motion from a random initial
condition. 

In order to generate non-trivial but easily understood dynamics, we
prepare a fifth particle in a classical-like Gaussian wavepacket $g(x,s)$ given
by
\begin{equation*}
  g(x,s) = C \exp\left[-(x-x_0)^2/(4\sigma^2) + \rmi k_0 x \right] \chi_{+1/2}(s), 
\end{equation*}
where $C$ is a normalization constant, $x_0$ the ``starting position''
of the particle, and $k_0$ is the ``starting momentum''. The parameter
$\sigma$ controls the width of the wavepacket. For our experiment, we
choose $x_0 = 10$, $k_0 = 1.2$, and, $\sigma = 1.25$. Note that we
(arbitrarily) choose spin $+1/2$ for this particle. The initial MCTDHF
state is
then
\begin{equation*}
  \ket{\Psi} = g^\dag \ket{\Psi_4},
\end{equation*}
where $g^\dag$ is the creation operator associated with $g(x,s)$. The
complete wavefunction therefore intuitively describes an incoming
electron on collision course with a bound beryllium-like ``atom'' in
the ground state.

In terms of the MCTDHF parameters, the addition of the particle simply
corresponds to extending the orbital matrix $\Phi$ with an extra column
(orthogonalized against the others), in addition to a lot of zeroes in
the MCTDHF coefficient vector $A$.

\begin{figure}
  \begin{center}
    \includegraphics{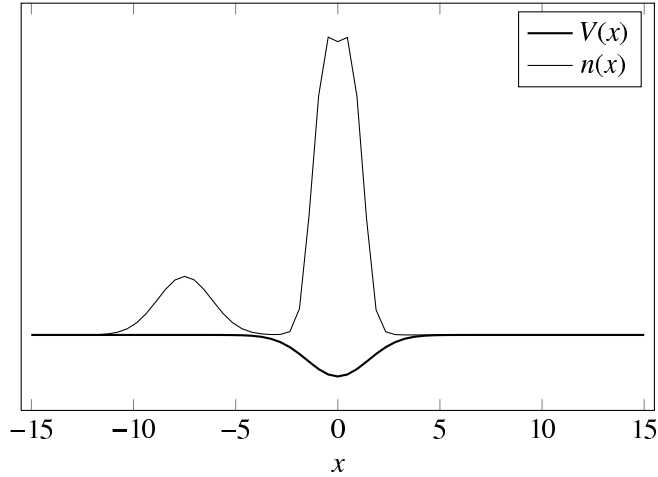}
  \end{center}
  \caption{Particle density $n(x)$ of initial wavefunction and
    Gaussian well $V(x)$. Units are arbitrary.\label{fig:ground-state}}
\end{figure}

The spatial particle density $n(x) = n(x,-1/2) + n(x,+1/2)$ of $\ket{\Psi}$ is shown
in Fig.~\ref{fig:ground-state} along with the confining potential. The
density $n(x,s)$ is given by the diagonal of the reduced one-body density
matrix $\gamma(x,s,x',s')$, viz,
\begin{equation*}
  \gamma({x,s,x',s'}) = \braket{\Psi|\boldpsi^\dag(x,s)\boldpsi(x',s')|\Psi},
\end{equation*}
or in terms of the coefficients $\rho^{(1)} = [\rho^q_p]$ and the
orbitals,
\begin{equation}
  \gamma = \Phi \rho^{(1)} \Phi^H.
  \label{eq:densmat}
\end{equation}

Having obtained the MCTDHF initial wavefunction, the OATDCCD initial
condition is computed as follows. We first transform the orthonormal
orbitals $\Phi$ to generate orthonormal so-called Brueckner orbitals
$\Phi_\text{B}$. The wavefunction amplitudes are transformed
accordingly. By definition \cite{Lowdin1962}, the Brueckner orbitals
optimize the overlap with the reference determinant, which is actually
equivalent to the convenient property that the singles amplitudes vanish
identically, the so-called Brillouin--Brueckner theorem. We now have
\begin{align}
  \ket{\Psi} = \braket{\phi_\text{B}|\Psi}(1 + A_2 + A_3 +
  \cdots)\ket{\phi_\text{B}} = \braket{\phi_\text{B}|\Psi} e^{T_2 + T_3 +
    \cdots} \ket{\phi_\text{B}},
  \label{eq:initial}
\end{align}
where $\ket{\phi_\text{B}}$ is the determinant that has maximum
overlap with $\ket{\Psi}$. We now simply take the OATDCC orbitals to
be $\Phi_\text{B}$ (and $\tPhi=\Phi_\text{B}^H$), and let $T_2$ be as
in Eqn.~(\ref{eq:initial}), i.e., we perform a projection in CC
amplitude space.

It remains to define $\Lambda_2$. We note that for the MCTDHF limit of
OATDCC, $\bra{\tilde{\Psi}} = \bra{\Psi}/\braket{\Psi|\Psi}$, which gives
$\Lambda_n$ such that
\begin{equation*}
  \bra{\tilde{\Psi}} e^{T_2^\dag + T_3^\dag + \cdots} =
    \bra{\phi_\text{B}}(1 + \Lambda_1 + \Lambda_2 + \cdots),
\end{equation*}
which is used to extract $\Lambda_2$ algebraically in the computer
code.

We comment that the initial condition computed in this way only is a
critical point of the coupled cluster energy to within an
approximation, albeit a very good one. Alternatively, we could solve
the CCD equations in the Brueckner basis, which gives very similar results.

We comment that the  OATDCC counterpart of the density matrix (\ref{eq:densmat}) is
\begin{equation}
  \gamma = \Phi \rho^{(1)} \tPhi.
  \label{eq:densmat-cc}
\end{equation}
The density plot of the OATDCC initial condition is visually
indistinguishable from the MCTDHF initial condition, so we do not plot
it separately in Fig.~\ref{fig:ground-state}.

\subsection{Results}
\label{sec:experiment-results}

\begin{figure}
  \begin{center}
    \includegraphics{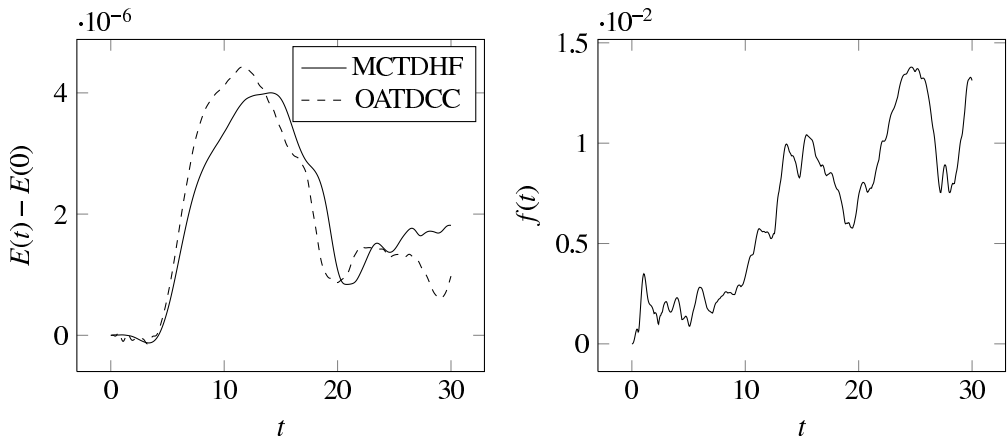}
  \end{center}
  \caption{(Left) Energy conservation in the MCTDHF and OATDCCD schemes for
    the numerical experiment. The deviation from the initial energy is
  shown. The initial energies are $E_\text{MCTDHF}(0) =
  -12.2102145$ and $E_\text{OATDCCD}(0) = -12.2104173$. The gap between
  these numbers are due to neglection of the triples amplitudes and
  higher in the ground state. (Right) Integral of imaginary part of
  particle density, see Eqn.~(\ref{eq:imag-integral}).\label{fig:energy-conservation}}
\end{figure}

Having obtained initial conditions, these are propagated in time with
$\Delta t = 0.005$ until $t=t_\text{final}=30$. The energy is
conserved to a very high precision, see
Fig.~\ref{fig:energy-conservation}. The energy-conservation also
improves with reduced time step, as is expected. 

As discussed in
Section~\ref{sec:bi-vp-3}, expectation values may gain small imaginary
parts. Along the CC computation we therefore monitor
\begin{align}
  f(t) = \sum_s \int |\Im n(x,s,t)| \rmd x.
  \label{eq:imag-integral}
\end{align}
In Fig.~\ref{fig:energy-conservation} $f(t)$ is displayed, and it is
indeed a small number compared to the particle number $N=5$.

\begin{figure}
  \begin{center}
    \includegraphics{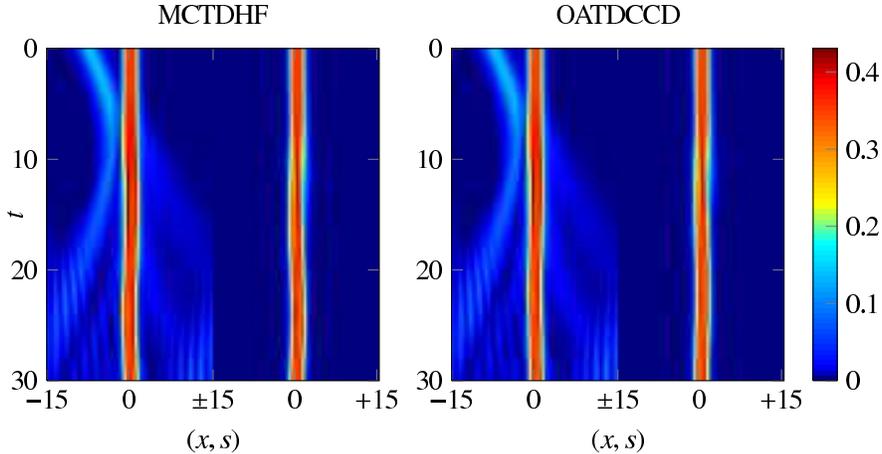}
  \end{center}
  \caption{Electron density plot as function of time for each
    calculation. MCTDHF is on the left, while OATDCC is on the
    right. The $x$-axis is divided into spin up (left half) and spin
    down (right half). The two densities are seen to be very similar,
    see also Fig.~\ref{fig:densplot-errors}. The incident electron is
    clearly reflected and  
    partially transmitted through the initially stationary beryllium
    atom. After the collision, the atom is seen to exhibit
    oscillations. The interference fringes at the end of the
    simulations are due to boundary effects.
    \label{fig:densplot}}
\end{figure}

In Fig.~\ref{fig:densplot} we show the density as function of $t$ of
each calculation side by side. They are seen to agree
qualitatively. The density evolution clearly shows how the incident
electron interacts with the beryllium ``atom''. Some of the density
is clearly transmitted and reflected from the atom, while the atom is
slightly perturbed, performing small amplitude oscillations. This
demonstrates that manybody effects in the simulation are significant,
and that the OATDCCD calculation captures these well.

\begin{figure}
  \begin{center}
    \includegraphics{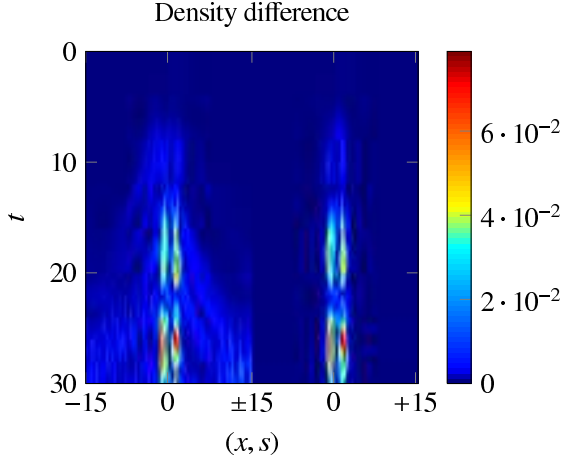}
  \end{center}
  \caption{Differences between the OATDCC and MCTDHF electron
    densities calculated. The main feature is a phase-shift in the
    atom oscillations.\label{fig:densplot-errors}}
\end{figure}

Quantitatively, the densities show some differences after the
collision event that may look like a phase shift 
in the oscillation of the atom part. We have not investigated further,
but conjecture this to be a result of the fact that the CC
approximation changes the spectrum slightly. The absolute value of the
density difference is shown in 
Fig.~\ref{fig:densplot-errors}. 

\section{Conclusion}
\label{sec:conclusion}

The bivariational principle for Schrödinger dynamics has been
discussed at length, and the orbital-adaptive time-dependent
coupled-cluster method (OATDCC) was developed. The method can be
viewed as a systematic hierarchy of approximations to the highly
successful multiconfigurational time-dependent Hartree method for
fermions (MCTDHF), with simple time-dependent Hartree--Fock as the
simplest case. The doubles approximation (OATDCCD) was discussed in
detail, and numerical experiment performed showing that the method
gives sensible results.  OATDCC scales polynomially where MCTDHF
scales exponentially with the number of particles $N$.

It was observed that imaginary time propagation for eigenvalue
computation does not seem feasible for the OATDCC method. Studying
methods for solving the time-independent orbital-adaptive CC
should be a useful line of research. Such an eigenvalue computation
method would constitute a hierarchy of approximations to the
multiconfigurational Hartree--Fock method.

OATDCC is easily generalized to bosonic systems, with particularly
interesting applications to Bose--Einstein condensates (BEC). The
resulting method will approximate the MCTDH method for bosons
\cite{Alon2008}, and can possibly treat substantially more
particles. In fact, the CC approximation is very well suited to
describe a BEC since it naturally captures the idea of excitations on
top of a condensate.

\subsection*{Acknowledgments}

 The author wishes to thank Prof.~Christian Lubich of Universität
 Tübingen, Germany, for fruitful discussions and Prof.~Lars Bojer
 Madsen of Aarhus Universitet, Denmark, for constructive feedback on the
 manuscript.  This work is supported by the DFG priority programme
 SPP-1324 \cite{SPP-1324}. Further financial support of CMA, University of Oslo, is
 gratefully acknowledged.

\appendix

\section{Algebraic expressions for CCD}
\label{sec:algebraic}

Here we list algebraic expressions for various quantities appearing in
the OATDCC method using a doubles only ansatz, i.e., CCD. The
expressions are computed using the second quantization toolbox in the
\textsc{Python} library \textsc{SymPy} \cite{SymPy2009}.

The expression for the expectation functional becomes
\begin{align}
  \mathcal{E}_H[\lambda,\tau,\tPhi,\Phi] &= \braket{\tphi|He^T|\phi} +
  \sum_\mu \lambda_\mu \pdiff{}{\lambda_\mu}
  \mathcal{E}_H[\lambda,\tau,\tPhi,\Phi] \label{eq:exp-onebody} \\
&= \braket{\tphi|He^T|\phi} +
  \sum_\mu \lambda_\mu \braket{\tphi_\mu|e^{-T}H e^T|\phi} \\
&=  \braket{\tphi|He^T|\phi} +
  \frac{1}{4}\sum_{ijab} \lambda_{ab}^{ij} \braket{\tphi_{ij}^{ab}|e^{-T}H
    e^T|\phi} \label{eq:exp-twobody}
\end{align}
where we have used linearity of $\mathcal{E}_H$ in $\lambda_\mu$, and
where the latter expression explicitly states the expansion in the
CCD case.

To obtain computational formulae, we consider separately the one and
two-body terms in $H$, i.e.,
\begin{equation}
  H^{(1)} = \sum_{pq} h^p_q c^\dag_p \tc_q, \quad   h^p_q =
  \braket{\tvphi_p|H^{(1)}|\vphi_q}
\end{equation}
and
\begin{equation}
  H^{(2)} = \frac{1}{4}\sum_{pqrs} u^{pr}_{qs} c^\dag_p c^\dag_r \tc_s
  \tc_q, \quad
  u^{pr}_{qs} = \braket{\tvphi_p\tvphi_r|H^{(2)}|\vphi_q\vphi_s}_{AS}
\end{equation}
The coefficients $u^{pr}_{qs}$ are the the anti-symmetrized two-body
integrals, see Equations~(\ref{eq:two-body-integral-1}) and
(\ref{eq:two-body-integral-2}). We get
\begin{subequations}
\label{eq:ccd-rgy}
\begin{align}
  \braket{\tphi|H^{(1)}e^{T}|\phi} &=  h^{i}_{i} \\
  \braket{\tphi|H^{(2)}e^{T}|\phi} &= \frac{1}{4} \tau^{ab}_{ij}
  u^{ij}_{ab} +
  \frac{1}{2} u^{ij}_{ij} 
\end{align}
\end{subequations}
for the CCD energy, and
\begin{subequations}
\label{eq:lambda-derivatives}
\begin{align}
  \pdiff{}{\lambda_{ab}^{ij}}\mathcal{E}_{H^{(1)}} &= - h^{a}_{c}
  \tau^{bc}_{ij} P(ab) + h^{k}_{i} \tau^{ab}_{jk} P(ij)  \\
  \pdiff{}{\lambda_{ab}^{ij}}\mathcal{E}_{H^{(2)}} &= - \tau^{ab}_{ik}
  u^{kl}_{jl} P(ij) + \frac{1}{2} \tau^{ab}_{il} \tau^{dc}_{jk} u^{kl}_{dc}
  P(ij) + \frac{1}{4} \tau^{ab}_{kl} \tau^{dc}_{ij} u^{kl}_{dc} +
  \frac{1}{2} \tau^{ab}_{kl} u^{kl}_{ij} + \frac{1}{2} \tau^{ac}_{ij}
  \tau^{bd}_{kl} u^{kl}_{dc} P(ab) \notag\\ &+ \tau^{ac}_{ij} u^{bk}_{ck} P(ab) -
  \tau^{ac}_{ik} \tau^{bd}_{jl} u^{kl}_{dc} P(ab) + \tau^{ac}_{ik} u^{bk}_{jc}
  P(ab) P(ij) + \frac{1}{2} \tau^{dc}_{ij} u^{ab}_{dc} + u^{ab}_{ij}
\end{align}
\end{subequations}
for the derivatives with respect to $\lambda_{ab}^{ij}$.
Inserting these expressions back into Eqn.~(\ref{eq:exp-onebody}), the
complete expression for the CCD expectation value functional becomes
\begin{align}
  \mathcal{E}_H[\lambda,\tau,\tPhi,\Phi] &= \frac{1}{2} h^{a}_{b} \lambda^{ij}_{ac} \tau^{bc}_{ij} +
  h^{i}_{i} - \frac{1}{2} h^{j}_{i} \lambda^{ki}_{ab} \tau^{ab}_{kj} -
  \frac{1}{2} \lambda^{ij}_{ab} \tau^{ab}_{ki} u^{kl}_{lj} + \frac{1}{8}
  \lambda^{ij}_{ab} \tau^{ab}_{kj} \tau^{dc}_{li} u^{kl}_{dc} \\ & + \frac{1}{16}
  \lambda^{ij}_{ab} \tau^{ab}_{kl} \tau^{dc}_{ij} u^{kl}_{dc} + \frac{1}{8}
  \lambda^{ij}_{ab} \tau^{ab}_{kl} u^{kl}_{ij} + \frac{1}{8} \lambda^{ij}_{ab}
  \tau^{ab}_{li} \tau^{dc}_{kj} u^{kl}_{dc} + \frac{1}{2} \lambda^{ij}_{ab}
  \tau^{ac}_{ij} u^{bk}_{ck} \\&+ \lambda^{ij}_{ab} \tau^{ac}_{ki} u^{bk}_{cj} -
  \frac{1}{2} \lambda^{ij}_{ab} \tau^{ac}_{kj} \tau^{db}_{li} u^{kl}_{dc} -
  \frac{1}{4} \lambda^{ij}_{ab} \tau^{ac}_{kl} \tau^{db}_{ij} u^{kl}_{dc} \\&+
  \frac{1}{8} \lambda^{ij}_{ab} \tau^{dc}_{ij} u^{ab}_{dc} + \frac{1}{4}
  \lambda^{ij}_{ab} u^{ab}_{ij} + \frac{1}{4} \tau^{ab}_{ij} u^{ij}_{ab} +
  \frac{1}{2} u^{ij}_{ij} 
\end{align}
To solve the equations of motion for the amplitudes
$\lambda_{ab}^{ij}$ we also need the derivatives of $\mathcal{E}_H$
with respect to $\tau_{ij}^{ab}$. For the one-body part
\begin{subequations}
\label{eq:tau-derivatives}
\begin{equation}
  \pdiff{}{\tau_{ij}^{ab}}\mathcal{E}_{H^{(1)}} = h^{i}_{k}
  \lambda^{jk}_{ab} P(ij) - h^{c}_{a} \lambda^{ij}_{bc} P(ab)  ,
\end{equation}
and for the two-body part,
\begin{align}
  \pdiff{}{\tau_{ij}^{ab}}\mathcal{E}_{H^{(2)}} &=  - \frac{1}{2}
  \lambda^{ij}_{bc} \tau^{dc}_{kl} u^{kl}_{ad} P(ab) - \lambda^{ij}_{bc} u^{ck}_{ak}
  P(ab) + \frac{1}{4} \lambda^{ij}_{dc} \tau^{dc}_{kl} u^{kl}_{ab}
  \notag \\&+
  \frac{1}{2} \lambda^{ij}_{dc} u^{dc}_{ab} + \frac{1}{2} \lambda^{jk}_{ab}
  \tau^{dc}_{kl} u^{il}_{dc} P(ij) + \lambda^{jk}_{ab} u^{il}_{kl} P(ij) -
  \lambda^{jk}_{bc} \tau^{dc}_{kl} u^{il}_{ad} P(ab) P(ij) \notag \\&+ \lambda^{jk}_{bc}
  u^{ic}_{ak} P(ab) P(ij) + \frac{1}{2} \lambda^{jk}_{dc} \tau^{dc}_{kl}
  u^{il}_{ab} P(ij) + \frac{1}{4} \lambda^{kl}_{ab} \tau^{dc}_{kl} u^{ij}_{dc}
  + \frac{1}{2} \lambda^{kl}_{ab} u^{ij}_{kl} \notag \\&- \frac{1}{2} \lambda^{kl}_{bc}
  \tau^{dc}_{kl} u^{ij}_{ad} P(ab) + u^{ij}_{ab} 
\end{align}
\end{subequations}
The operator $P(ij)$ is an
anti-symmetrizer: $f(ij)P(ij)=f(ij)-f(ji)$, and similarly for
$P(ab)$. The appearance of a $P(ij)$ or a $P(ab)$ should be ignored
for the invocation of the summation convention.

Our calculations for expectation values and derivatives are of course
valid for any one- or two-body operator.

Expressions for the reduced density matrices $\rho^q_p$ are readily obtained from 
$\mathcal{E}_{H^{(1)}}$ by using $h^{p'}_{q'} = \delta^{p'}_p
\delta_{q'}^q$. We get $\rho^a_i=\rho^i_a=0$, while
\begin{subequations}
  \label{eq:rho-1}
  \begin{align}
    \rho^{j}_i &= \mathcal{E}_{c^\dag_i \tc_j} = \delta^k_i \delta_k^j -
    \frac{1}{2}\delta^l_i\delta^j_m \lambda^{kl}_{ab}\tau^{ab}_{km} =
    \delta^j_i - \frac{1}{2}\lambda^{kj}_{ab}\tau^{ab}_{ki} \label{eq:rho-occ}\\
    \rho^{b}_a &= \mathcal{E}_{c^\dag_a \tc_b} =
    \frac{1}{2}\delta^d_a\delta^b_e \lambda^{ij}_{dc}\tau^{ec}_{ij} =
    \frac{1}{2}\lambda^{ij}_{ac}\tau^{bc}_{ij} \label{eq:rho-vir}
  \end{align}
\end{subequations}
Finally, we compute the two-body reduced density matrix. We only list nonzero
elements.
\begin{subequations}
\begin{align}
  \rho^{kl}_{ij} &= P(ij)\delta^k_i \delta^l_j - P(ij)P(kl)\frac{1}{2}
  \delta^k_i \lambda^{lm}_{cd} \tau^{cd}_{jm} + \frac{1}{2} \lambda^{kl}_{cd} \tau^{cd}_{ij} \\
%\rho^{KD}_{IJ} &= 0 \\
%\rho^{CL}_{IJ} &= 0 \\
\rho^{ab}_{ij} &= 
- P(ab)\frac{1}{2} \lambda^{kl}_{cd} \tau^{ac}_{ij} \tau^{bd}_{kl} + P(ij)\lambda^{kl}_{cd}
\tau^{ac}_{ik} \tau^{bd}_{jl}  \notag \\
&\qquad + P(ij)\frac{1}{2}
\lambda^{kl}_{cd} \tau^{ab}_{il} \tau^{cd}_{jk}  + \frac{1}{4} \lambda^{kl}_{cd} \tau^{ab}_{kl}
\tau^{cd}_{ij} + \tau^{ab}_{ij} \\
%\rho^{KL}_{IB} &= 0 \\
\rho^{jb}_{ia} &= -\rho^{bj}_{ia} = -\rho^{jb}_{ai} = \rho^{bj}_{ai} =
\frac{1}{2} \delta^j_i \lambda^{kl}_{ac} \tau^{bc}_{kl} - \lambda^{jk}_{ac}
\tau^{bc}_{ik} \\
%\rho^{CL}_{IB} &=
%-\frac{1}{2} \delta_{IL} \lambda^{ij}_{Ba} \tau^{Ca}_{ij} + \lambda^{Li}_{Ba}
%\tau^{Ca}_{Ii} \\
%\rho^{CD}_{IB} &= 0 \\
%\rho^{KL}_{AJ} &= 0 \\
%\rho^{KD}_{AJ} &= -\frac{1}{2} \delta_{JK} \lambda^{ij}_{Aa} \tau^{Da}_{ij} +
%\lambda^{Ki}_{Aa} \tau^{Da}_{Ji} \\
%\rho^{CL}_{AJ} &=  \frac{1}{2} \delta_{JL} \lambda^{ij}_{Aa} \tau^{Ca}_{ij} -
%\lambda^{Li}_{Aa} \tau^{Ca}_{Ji} \\
%\rho^{CD}_{AJ} &= 0 \\
\rho^{ij}_{ab} &=  \lambda^{ij}_{ab} \\
%\rho^{KD}_{AB} &= 0 \\
%\rho^{CL}_{AB} &= 0 \\
\rho^{cd}_{ab} &= \frac{1}{2} \lambda^{ij}_{ab} \tau^{cd}_{ij}
\end{align}
\end{subequations}

\section{Derivation of MCTDHF}
\label{sec:mctdhf-derivation}

The purpose of this section is to provide a brief derivation of the MCTDHF
method independently of the existing derivations in the literature,
using the time-dependent bivariational principle.

The bivariational MCTDHF manifold is
\begin{equation}
  \mathcal{M}_\text{MCTDHF} = \bigcup_{\tPhi\Phi=I}
  \tilde{\mathcal{V}}[\tPhi]\times \mathcal{V}[\Phi],
\end{equation}
where the union is taken over all biorthogonal choices of
orbitals. That is to say, $\ket{\Psi}$ is an arbitrary vector in the
discrete (FCI) Hilbert space generated by $\Phi$, and $\bra{\tPsi}$ is
an arbitrary vector in the discrete space generated by the dual orbitals
$\tPhi$:
\begin{align}
  \ket{\Psi} &= \sum_{\mu} A^\mu \ket{\phi_\mu} \\
  \bra{\tPsi} &= \sum_{\mu} \tilde{A}_\mu \bra{\tphi^\mu} ,
\end{align}
where we include the reference determinants in the sums by defining
$\ket{\phi_0}\equiv \ket{\phi}$ and $\bra{\tphi_0}\equiv \bra{\tphi}$,
i.e., $\mu=0$ is the reference. Since the FCI spaces are linear it is
allowed to remove the phase and normalization ambiguity in
Eqn.~(\ref{eq:td-functional}) by requiring $\braket{\tPsi|\Psi}=1$.

The time derivative of $\ket{\Psi}$ is
\begin{equation}
  \pdiff{}{t}\ket{\Psi} = \sum_\mu \dot{A}^\mu \ket{\phi_\mu} + D \ket{\Psi},
\end{equation}
where $D$ is defined in Eqn.~(\ref{eq:D-def-and-psi-derivative}). We
obtain two expressions for the action functional, which are equivalent:
\begin{align}
  \mathcal{S}[\tilde{A},A,\tPhi,\Phi] &= \int_{0}^{T} \rmi\hbar
  \tilde{A}_\mu \dot{A}^\mu + \rmi\hbar\tilde{A}_\mu D^\mu_\nu A^\nu -
  \tilde{A}_\mu H^\mu_\nu A^\nu \; \rmd t \\
&= \int_{0}^{T} \rmi\hbar \tilde{A}_\mu \dot{A}^\mu  + \rho^{q}_p (\rmi\hbar \eta^p_q - h^p_q) - \frac{1}{4}
\rho^{qs}_{pr} u^{pr}_{qs} \; \rmd t,
\end{align}
where
\begin{equation}
  D^{\mu}_{\nu} \equiv \braket{\tphi^\mu|D|\phi_\nu}, \quad H^{\mu}_{\nu}
  \equiv \braket{\tphi^\mu|H||\phi_\nu}, \quad \eta^p_q \equiv \braket{\tvphi_p|\dot{\vphi}_q},
\end{equation}
and where
\begin{equation}
  \rho^q_p \equiv \braket{\tPsi|c^\dag_p \tc_q|\Psi},\quad
  \rho^{qs}_{pr} \equiv \braket{\tPsi|c^\dag_p c^\dag_r \tc_s \tc_q|\Psi}
\end{equation}
are the reduced one- and two-body density matrices.

In a similar way as for the CC ansatz, it can be shown that mixing of
the orbitals $\Phi\rightarrow \Phi G$ and $\tPhi \rightarrow G^{-1}
\tPhi$, $G$ being an arbitrary invertible $L\times L$ matrix, with a
corresponding inverse operation on $A$ and $\tilde{A}$ leaves the
action invariant (except for a total time derivative with vanishing
variation). This implies that \emph{all} the inner products $\eta^p_q$
can be chosen arbitrarily, and we choose $\eta^p_q\equiv 0$ for
simplicity, which implies $D^\mu_\nu\equiv 0$. This corresponds to the
usual gauge choice in MCTDHF theory, see Ref.~\cite{Beck2000}. The
resulting permissible variations in $\vphi_p$ are then of the form
$\delta \vphi_p = \chi = Q \chi$, with $Q = 1 - \Phi\tPhi$. Similarly,
the permissible variations in $\tvphi_p$ are $\delta\tvphi_p =
\tilde{\chi} = \tilde{\chi}Q$. The $P$-space variations are
identically zero in this gauge.

Performing first the variations in $\tilde{A}_\nu$ and $A^\mu$,
respectively, we obtain  the
equations
\begin{equation}
 \rmi\hbar \dot{A}^\nu = H^\nu_\mu A^\mu \quad \text{and} \quad
 -\rmi\hbar \dot{\tilde{A}}_\mu = \tilde{A}_\nu H^\nu_\mu ,
\end{equation}
i.e., the time-dependent Schrödinger equation in a moving basis and
its dual equation. Since $H^\nu_\mu = \braket{\tphi^\nu|\Pi H
  \Pi|\phi_\mu}$, the operator expression (\ref{eq:projected-h}) is
useful for implementations.

Performing the variations in $\tvphi_p$ we obtain, in a manner similar
to Section~\ref{sec:eom-deriv}, the equation
\begin{equation}
  \rmi\hbar \sum_q \rho^q_p \ket{\dot{\vphi}_q} = Q \left[ \sum_q
    \rho^q_p h \ket{\vphi_q} + \sum_{qrs} \rho^{qs}_{pr} W^r_s
    \ket{\vphi_q} \right],
  \label{eq:mctdh-orbital-eom}
\end{equation}
and performing the variation in $\vphi_q$ we obtain
\begin{equation}
  -\rmi\hbar \sum_p \rho^q_p \bra{\dot{\tvphi}_p} =  \left[ \sum_p
    \rho^q_p  \bra{\tvphi_p}h + \sum_{prs} \rho^{qs}_{pr} 
    \bra{\vphi_p}W^r_s \right] Q,
  \label{eq:mctdh-orbital-eom-2}
\end{equation}
which are identical in form to Eqns.~(\ref{eq:eom-5}) and
(\ref{eq:eom-6}). 

Assume now that $\tPhi(0) = \Phi(0)^H$ at time $t=0$, and that
$\tilde{A}_\mu(0) = A^\mu(0)^*$. Then clearly $\rho^q_p =
(\rho_q^p)^*$ and $\rho^{qs}_{pr}=(\rho^{pr}_{qs})^*$. Similarly,
$h^p_q = (h^q_p)^*$ and $u^{pr}_{qs} = (u^{qs}_{pr})^*$.

From this it follows that $\bra{\dot{\tvphi}_p} =
\ket{\dot{\vphi}_p}^\dag$, $\dot{\tilde{A}}_\mu = (\dot{A}^\mu)^*$ and
hence $\bra{\tPsi(t)} = \ket{\Psi(t)}^\dag$ for all $t$.

Comparing with the equations of motion in, say,
Ref.~\cite{Zanghellini2003}, we see that we indeed have arrived at the
MCTDHF equations of motion.

\section{Some techinical proofs}
\label{sec:technical}

\subsection{Orbital transformations}

We consider the orbital transformation
\begin{equation}
  \Phi \larrow \Phi' = \Phi G, \quad \tilde{\Phi} \larrow
  \tilde{\Phi}' = G^{-1}\tilde{\Phi}
  \label{eq:rotation-app},
\end{equation}
where $G=\exp(g)$ is an $L\times L$ invertible matrix. (Recall that
any invertible matrix has a logarithm.) The transformation
preserves biorthogonality. The transformation is equivalent to transforming the
creation and annihilation operators as
\begin{equation}
  c^\dag_p \larrow \sum_q c^\dag_q G_{qp}, \quad   \tilde{c}_p \larrow
  \sum_q G^{-1}_{pq}  \tilde{c}_q.
  \label{eq:rotation-2-app}
\end{equation}
We wish to show, that Eqn.~(\ref{eq:rotation-2-app}) is again
equivalent to 
\begin{equation}
  c^\dag_p \larrow e^{\hat{g}} c^\dag_p e^{-\hat{g}}, \quad 
  \tc_q \larrow e^{\hat{g}} \tc_q e^{-\hat{g}}, \label{eq:rotation-3-app}
\end{equation}
where
\begin{equation}
  \hat{g} = \sum_{pq} g_{qp} c^\dag_q \tc_p.
  \label{eq:G-onebody-app}
\end{equation}
We begin by noting that the similarity transform can be expanded using
the BCH formula as
\begin{equation}
  e^{\hat{g}} x e^{-\hat{g}} = x + [\hat{g},x] +
  \frac{1}{2!}[\hat{g},[\hat{g},x]] + \cdots = \sum_{n=0}^\infty
  \frac{1}{n!}[\hat{g},x]_n
\end{equation}
where $[A,B]_n$ is the $n$-fold nested commutator. Suppose we can
show, that for all $n\geq 0$,
\begin{subequations}
\label{eq:proveme}
\begin{align}
  [\hat{g}, c^\dag_p]_n &= \sum_q (g^n)_{qp}
  c^\dag_q, \label{eq:proveme-a} \intertext{and}
  [\hat{g}, \tc_q]_n &= \sum_p (-1)^n(g^n)_{qp} \tc_p.\label{eq:proveme-b}
\end{align}
\end{subequations}
This would imply
\begin{align}
e^{\hat{g}} c^\dag_p e^{-\hat{g}} &= \sum_{n=0}^\infty \frac{1}{n!}
\sum_q (g^n)_{qp} c^\dag_q = \sum_q G_{qp} c^\dag_q \intertext{and}
e^{\hat{g}} \tc_q e^{-\hat{g}} &= \sum_{n=0}^\infty \frac{(-1)^n}{n!}
\sum_p (g^n)_{qp} \tc_p = \sum_p G^{-1}_{qp} \tc_p,
\end{align}
which would prove the result.

We prove Eqn.~(\ref{eq:proveme}) by induction, and for simplicity we
consider only (\ref{eq:proveme-a}). The proof for (\ref{eq:proveme-b})
is similar. For $n=0$, $(g^0)_{qp}=\delta_{qp}$, which shows that
$n=0$ is trivially true. Suppose now, that Eqn.~(\ref{eq:proveme-a})
holds for some $n\geq 0$. We show that it holds for $n+1$ as well. We get
\begin{equation}
[\hat{g},c^\dag_q]_{n+1} = [\hat{g},[\hat{g},c^\dag_q]_n] = [\hat{g},
\sum_{r}g^n_{rq} c^\dag_r] = \sum_{psr} g_{sp} g^n_{rq} [c^\dag_s
\tc_p, c^\dag_r] = \sum_{psr} g_{sp} g^n_{rq} \delta_{pr} c^\dag_s =
\sum_{s} g^{n+1}_{sp} c^\dag_s,
\end{equation}
which proves the induction step. We have used that 
\begin{equation}
  [c^\dag_s \tc_p, c^\dag_r] = \delta_{pr} c^\dag_s,
\end{equation}
which is easily calculated using the fundamental anticommutator.

\subsection{Amplitude equations}

In this section, we consider the derivation of the amplitude
equations~(\ref{eq:amp-eq}) from the variation of the OATDCC
functional with respect to the amplitudes $\lambda_\mu$ and
$\tau^\mu$, which are all independent variables. We therefore start
with varying a single amplitude $\lambda_\nu(t)$, i.e.,
$\delta\lambda_\mu(t) = 0$ for $\mu\neq \nu$, and $\delta\tau^\mu(t)=0$
for all $\mu\in\mathcal{I}$. Note that all variations vanish at the
end points $t=0$ and $t=T$ of the time interval. We wish to compute
\begin{equation}
  \delta \mathcal{S}[\lambda,\tau] = \delta \int_0^T \rmi\hbar\lambda_\mu(t)\dot{\tau}^\mu(t) -
  \mathcal{E}_{H-\rmi\hbar D_0}[\lambda(t),\tau(t)] \; \rmd t,
\end{equation}
where we have suppressed the dependence on $\tPhi$ and $\Phi$ in
$\mathcal{E}$ and $\mathcal{S}$ since they are held fixed in the variation. We obtain
\begin{align}
  \delta\mathcal{S}[\lambda,\tau] &= \int_0^T \rmi\hbar
  \delta\lambda_\nu(t)\dot{\tau}^\nu - \pdiff{\mathcal{E}_{H-\rmi\hbar
      D_0}[\lambda(t),\tau(t)]}{\lambda_\nu}\delta\lambda_\nu(t) \;
  \rmd t \\
 &= \int_0^T \delta\lambda_\nu(t)\left[\rmi\hbar\dot{\tau}^\nu -
   \pdiff{\mathcal{E}_{H-\rmi\hbar
       D_0}[\lambda(t),\tau(t)]}{\lambda_\nu}\right]\; \rmd t = 0.
\end{align}
Note that we do not invoke the summation convention.
Since the expression must vanish for \emph{all} choices of
$\lambda_\nu(t)$, it follows that 
\begin{equation}
  \rmi\hbar\dot{\tau}^\nu(t) =
  \pdiff{}{\lambda_\nu}\mathcal{E}_{H-\rmi\hbar D_0} [\lambda(t),\tau(t),\tPhi(t),\Phi(t)]
\end{equation}
must hold.

Similarly, we hold $\lambda_\mu$ fixed, and vary only a single
amplitude $\tau^\nu$.
\begin{align}
  \delta\mathcal{S}[\lambda,\tau] &= \int_0^T \rmi\hbar
  \lambda_\nu(t)\delta\dot{\tau}^\nu - \pdiff{\mathcal{E}_{H-\rmi\hbar
      D_0}[\lambda(t),\tau(t)]}{\tau^\nu}\delta\tau^\nu(t) \;
  \rmd t \\
 &= \int_0^T \delta\tau^\nu(t)\left[-\rmi\hbar\dot{\lambda}_\nu -
   \pdiff{\mathcal{E}_{H-\rmi\hbar
       D_0}[\lambda(t),\tau(t)]}{\tau^\nu}\right]\; \rmd t  +
 \rmi\hbar\left[\lambda_\nu(t)\delta\tau^\nu(t) \right]_{t=0}^T
\end{align}
where we have performed integration by parts. The boundary terms
vanish by definition of the variation, so that we get
\begin{equation}
  -\rmi\hbar\dot{\lambda}_\nu(t) =
  \pdiff{}{\tau^\nu}\mathcal{E}_{H-\rmi\hbar D_0} [\lambda(t),\tau(t),\tPhi(t),\Phi(t)].
\end{equation}

\bibliography{tdcc_refs.bib}

\end{document}